# Universal scaling laws of boundary-driven turbulence

Yong-Ying Zeng[1], Zi-Ju Liao[2], Jun-Yi Li[1,3], Wei-Dong Su[1,*]

[1]State Key Laboratory for Turbulence and Complex Systems and Department of Mechanics and Engineering Science, College of Engineering, Peking University; Beijing 100871, China

[2]Department of Mathematics, College of Information Science and Technology, Jinan University; Guangzhou 510632, China

[3]New Cornerstone Science Laboratory, Center for Combustion Energy, Key Laboratory for Thermal Science and Power Engineering of Ministry of Education, Department of Energy and Power Engineering, Tsinghua University; Beijing 100084, China

*Corresponding author. Email: swd@pku.edu.cn

**Abstract** Turbulence is a fundamental flow phenomenon, typically anisotropic at large scales and approximately isotropic at small scales. The classical Kolmogorov scaling laws (2/3, -5/3 and 4/5) have been well-established for turbulence without small-scale body forcing, describing second-order velocity structure functions, energy spectra, and third-order velocity structure functions in an intermediate small-scale range. However, their validity boundary remains unclear. Here, we identify new 1 and -2 scaling laws (replacing 2/3 and -5/3 laws) alongside the unchanged 4/5 law in the core region of boundary-driven turbulence, where energy is injected solely through viscous friction at moving boundaries. Local isotropy is recovered after high-pass filtering. Notably, odd-order velocity structure functions with and without absolute value exhibit distinct scaling exponents. A characteristic speed in the inertial range, derived from the constant ratio of third- to second-order structure functions, quantifies the time-averaged projectile speed at the bulk interface. Based on energy dissipation rate and the characteristic speed, a phenomenological framework for structure functions is developed together with a model for probability distributions of velocity increment at distinct small-scales. The universal scaling laws formulated can produce the full set of scaling exponents for low- and high-order velocity structure functions, including both the odd-orders' with and without absolute value, which are validated by direct numerical simulations and experimental datasets.

## Main:

Kolmogorov's phenomenological theory of turbulence, introduced in 1941[1–3], asserts that an arbitrary turbulent flow at a sufficiently high Reynolds number and distant from boundaries or singularities, is likely to exhibit local isotropy in statistics. In the inertial range, the second-order longitudinal velocity structure function follows the 2/3 power law, expressed as

$$S_2(\ell) \equiv \langle |\delta v_\ell|^2 \rangle = \langle \left| \left( \boldsymbol{u}(\boldsymbol{x} + \boldsymbol{\ell}, t) - \boldsymbol{u}(\boldsymbol{x}, t) \right) \cdot \boldsymbol{\ell}/\ell \right|^2 \rangle \sim \ell^{2/3}.$$

Here, $\boldsymbol{u}(\boldsymbol{x}, t)$ denotes the velocity at position $\boldsymbol{x}$ and time $t$, $\boldsymbol{\ell}$ is the displacement vector separating two points, and $\langle \cdot \rangle$ represents the statistical average. While higher-order structure functions occasionally deviate from the linear scaling law[4–10], the 2/3 law remains robust in overwhelming turbulent flows, even at moderate Reynolds numbers. Its spectral counterpart, the $-5/3$ scaling law for the energy spectrum $E(k) \sim k^{-5/3}$ (where $k \sim 1/\ell$ is the wavenumber in spectral space), has also become a standard law for fully developed turbulence[1,11,12]. In particular, homogeneous isotropic turbulence (HIT) in a periodic box, either without energy input or with a body force acting only at large scales, has become a benchmark for turbulence theories. However, this study, reports a novel observation: boundary-driven turbulence (BDT), where energy is injected via a steady tangentially moving boundary, exhibits fundamentally different scaling behavior in the small-scale range.

The flow field in a BDT consists of two distinct regions: the active boundary layers and the more passive bulk flow. The boundary layer is energized by viscous friction between the wall and the fluid, while the bulk is driven by the boundary layer. As the Reynolds number increases, plume structures form at the boundary, and large-scale flow structures develop in the bulk. For instance, in planar Couette flow (PCF), streamwise vortex rolls extend toward the wall as viscosity decreases[13–19]. Similarly, in Taylor-Couette flow (TCF), vortex rolls persist in the bulk region, even as the Reynolds number increases[9,10,20,21]. Large-scale circulation currents also appear in bulk flow of certain closed boundary flows, such as lid-driven cavity flows[22–24]. Local isotropy is difficult to achieve in the presence of such large-scale flow structures, even when the flow is not fully enclosed by boundaries. To isolate the effects of large-scale structures and restore isotropy, we apply a filtering method, allowing us to examine the scaling exponents using both the energy spectrum $E(k)$ and the second-order velocity structure function $S_2(\ell)$ of the filtered velocity field. Using helical wave decomposition (HWD) and direct numerical simulation (DNS), we explore BDT across different wall velocities and boundary geometries, including spherical, cylindrical, and channel domains. It should be emphasized that HWD, which is applicable to general three-dimensional domains, is an alternative to conventional Fourier analysis restricted to periodic boundary conditions.

**Table 1. Overview of boundary-driven turbulence cases investigated in the present study.** $\boldsymbol{u}_B$ denotes the velocity at the boundary. The unit vectors $\boldsymbol{e}_\theta$ and $\boldsymbol{e}_\varphi$ in spherical coordinates are aligned with the meridian and latitude, respectively. The unit vectors $\boldsymbol{e}_z$ and $\boldsymbol{e}_p$ correspond to the spin axis and precession axis in Cartesian coordinates, respectively. $\boldsymbol{e}_p$ can be expressed as $\boldsymbol{e}_p = \sin\alpha\,\boldsymbol{e}_y + \cos\alpha\,\boldsymbol{e}_z$, where $\alpha$ is the angel between spin and precession axes, set to 60 degrees in the cases studied. The Poincaré number, denoted by $\Gamma = \Omega_p/\Omega_s$, is fixed at $-0.3$. The vector $\boldsymbol{e}_x$ is the unit vector along the direction of motion of the upper planar plate.

| Description of turbulence case | | Acronyms | Boundary velocity formulation |
|---|---|---|---|
| Sphere | Driven by prescribed opposite steady rotation of the upper and lower hemispherical shells | SVK | $\boldsymbol{u}_B = \sin 2\theta\,\boldsymbol{e}_\varphi$ |
| | Driven by prescribed steady motion along meridians, similar to the boundary motion of the Hill's spherical vortex | SHF | $\boldsymbol{u}_B = \sin\theta\,\boldsymbol{e}_\theta$ |
| | Driven by prescribed a dipole-like steady motion on the surface, with opposite motion along the lines of latitude in the left and right hemispheres | SDF | $\boldsymbol{u}_B = \sin^2\theta\cos\varphi\,\boldsymbol{e}_\varphi$ |
| | Driven by precession of the sphere, i.e., constant rotation about one axis with constant orbital revolution along another axis[25] | SPF | $\boldsymbol{\Omega}_s = \Omega_s\boldsymbol{e}_z$ <br> $\boldsymbol{\Omega}_p = \Omega_p\boldsymbol{e}_p$ |
| Cylinder | Driven by prescribed rotation of the inner and outer cylinders with constant angular velocity, i.e., Taylor-Couette flow | TCF | Inner cylinder: $\Omega_i = 1$ <br> Outer cylinder: $\Omega_o = 0$ |
| Channel | Driven by opposite constant tangential motion of two parallel infinite planar plates | PCF | $\boldsymbol{u}_B = \boldsymbol{e}_x$ |

A summary of the BDT cases studied in the present work, along with their flow details, is provided in Table 1. We examine four cases with distinct boundary conditions within a spherical domain: SVK, SHF, SDF, and SPF. Additionally, we consider BDT configurations with partially open boundaries, such as TCF and PCF. In the SVK case, the azimuthal velocity on the spherical boundary is positive in the northern hemisphere and negative in the southern hemisphere, akin to the classical von Kármán flow driven by two infinite disks. The SHF case features a boundary velocity similar to Hill's spherical vortex[26] though the flow inside the sphere becomes turbulent at high Reynolds numbers, contrasting Hill's steady laminar flow. In the SDF case, the velocity distribution is symmetric about the prime meridian plane, with negative azimuthal velocity in the western hemisphere and positive velocity in the eastern hemisphere. In SPF, the orientation of the sphere's rotational axis changes over time.

DNS simulations across various Reynolds numbers (see Methods) confirm that all BDT configurations eventually reach a statistically steady state, in which energy dissipation balances the energy input from the boundary's tangential motion. This steady-state condition is crucial for the validity of our analysis. The typical flow structures in the three flow domains are illustrated in Fig. 1. A thin boundary layer forms near the wall, while large-scale circulation currents or vortices develop in the bulk, exhibiting multiscale plume structures. The velocity amplitude is lower and more uniform in the bulk, but higher and more variable in the boundary layer. Given the discrete characteristics of the boundary layers and the comparably more similar behavior of the bulk flows across different BDT configurations—our analysis focuses primarily on the bulk flows.

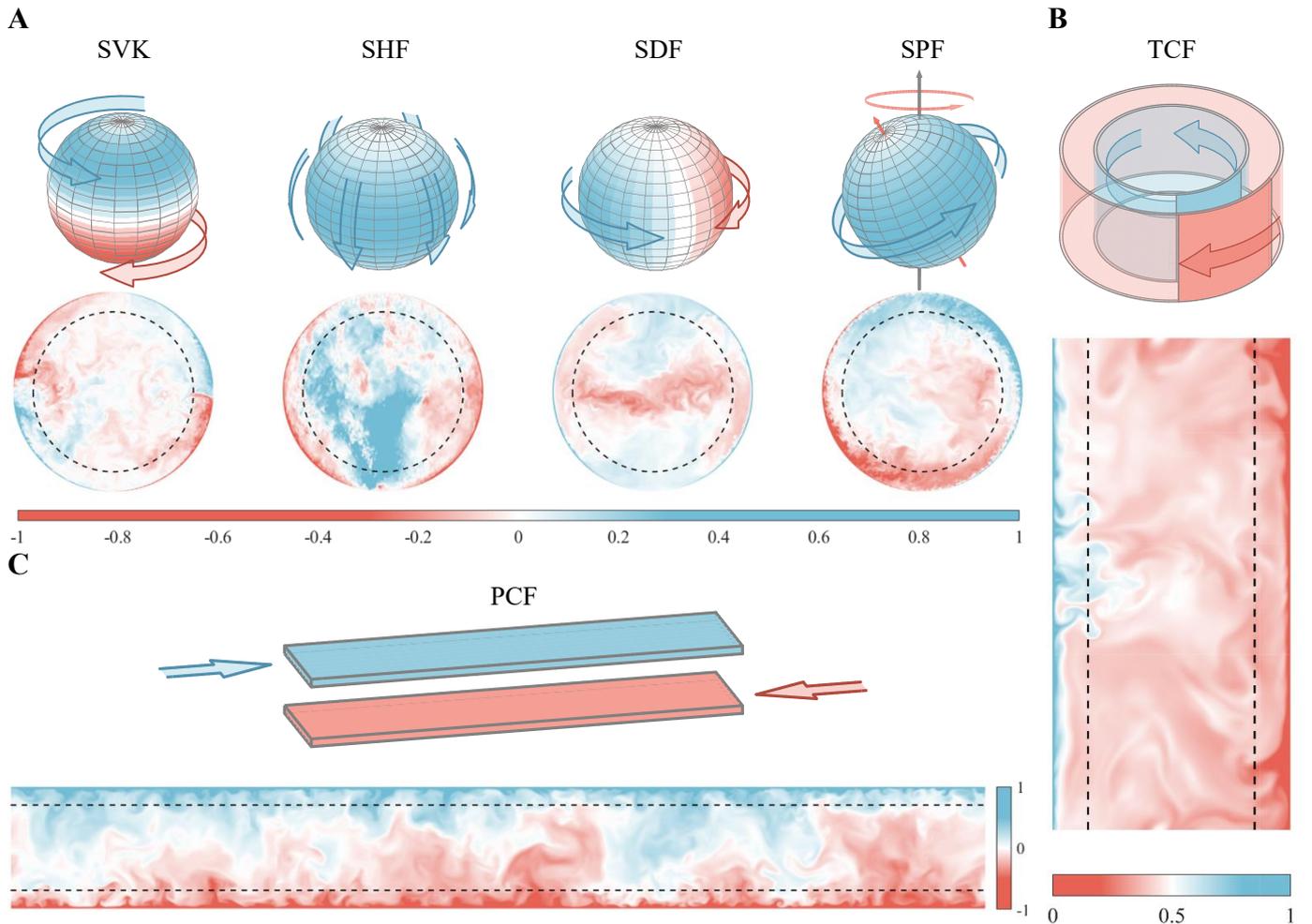

**Fig. 1. Flow visualizations of boundary-driven turbulence in various geometric domains.** In all panels, black dashed lines indicate the borders of the bulk region. **A, Spherical domains.** Velocity contours at $Re = 4.0 \times 10^4$ are shown in a meridian plane. The velocity components perpendicular to the meridional plane for the SVK, SDF, and SPF cases, and the velocity component normal to the equatorial plane for SHF, are depicted. The domain radius is 1, and the bulk flow radius is defined as 0.8. Multiple plume structures form in the boundary layer and are injected into the bulk as large circular currents. **B, Cylindrical domain.** The azimuthal velocity of the Taylor-Couette flow at $Ta = 4.62 \times 10^8$ and $Ro^{-1} = 0$ is presented. The bulk region is located at the center of the gap, with a width of 0.7 units, while the gap between the two coaxial cylinders is 1. Plume structures rise from the boundary layers and manifest as Taylor rolls. **C, Channel domain.** Streamwise velocity of PCF at $Re = 1.2 \times 10^4$ is illustrated. The distance between two planes is 1, with the bulk flow region having a width of 0.7. The bulk is populated with streamwise vortices, which extract energy from the boundary layer, resulting in a variety of plume structures emitted from these vortices.

# Scaling laws of energy spectra and second-order structure functions

In this section, we validate inertial-range scaling exponents in BDT using HWD. Global energy spectra $E(k)$ derived from HWD (Methods) for eleven cases spanning varied Reynolds numbers and boundary conditions are shown in Fig. 2. For all spherical-wall cases (SVK, SHF, SDF, SPF), we observe a power-law scaling of $E(k) \sim k^{-2}$, with the scaling range expanding to higher wavenumbers as the Reynolds number increases. The same scaling behavior is corroborated by Fourier analysis (Fig. S1) carried out in a cubic subregion of the spherical bulk. Similar scaling behavior is observed in both TCF and PCF.

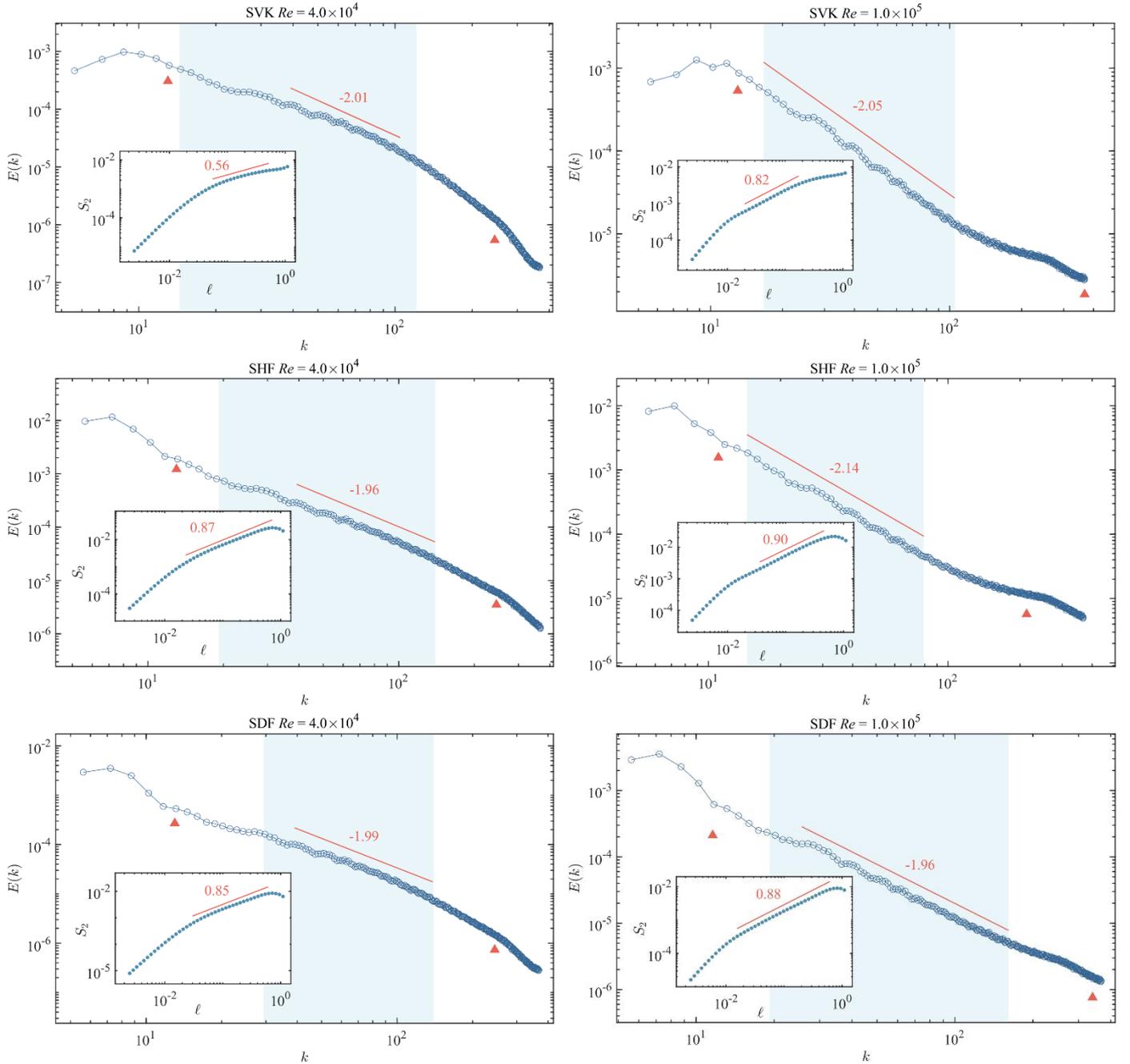

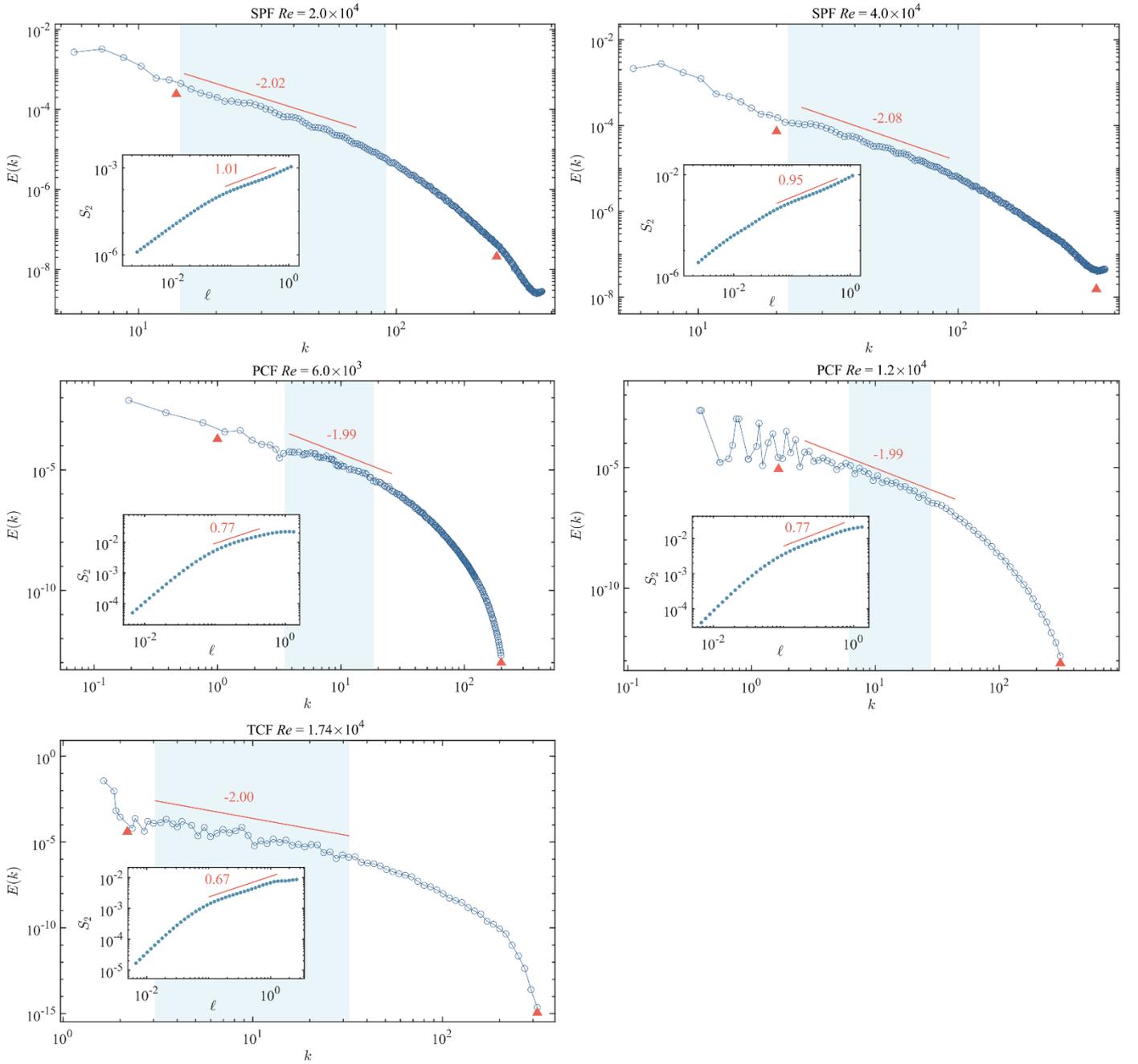

**Fig. 2. Spatial energy spectra and second-order velocity structure functions for BDT cases.** The time-averaged global energy spectrum $E(k)$ is presented as a function of wavenumber $k$. The definition of the wavenumber is based on the helical wave modes, which are dependent on the geometry of the flow domain. The inset in each panels shows the time-averaged spatial second-order velocity structure function $S_2(\ell)$ for the non-filtering (original) field. The energy spectrum follows a $E(k) \sim k^{-2}$ scaling law in all cases. The second-order structure function exhibits the following relationships: $S_2 \sim \ell^{0.82}$ for SVK, $S_2 \sim \ell^{0.90}$ for SHF, $S_2 \sim \ell^{0.88}$ for SDF, $S_2 \sim \ell^{0.95}$ for SPF, $S_2 \sim \ell^{0.77}$ for PCF, and $S_2 \sim \ell^{0.67}$ for TCF. The fitting range for $E(k)$ and $S_2(\ell)$ corresponds to the range marked by the red solid line in figures. In subsequent analysis, the velocity field is reconstructed using a filter, retaining only the helical modes located between the two red triangle symbols. The blue shaded region in wavenumber space corresponds to the scaling range of extended self-similarity for filtered velocity field.

The insets of Fig. 2 demonstrate the second-order structure functions $S_2(\ell)$ for the bulk flow in each case. In the SHF, SDF, and SPF cases, the scaling range becomes more pronounced and extends into the small-scale range as the Reynolds number increases, with exponents ranging from 0.8 to 1. In contrast, the scaling range for SVK and PCF cases at low Reynolds number is less clearly defined, though clear scaling ranges emerge as viscosity decreases, yielding exponents of 0.82 and 0.77, respectively. For TCF, the scaling exponent is found to be 0.67. In contrast to the energy spectrum, no consistent scaling exponent for $S_2(\ell)$ emerges across all cases. Traditional HIT predicts a direct relationship between the scaling exponents of the second-order velocity structure function $S_2(\ell)$ and the energy spectrum $E(k)$, which derived through dimensional analysis. According to Kolmogorov 1941 theory, this results in the scaling laws $S_2 \sim \ell^{2/3}$ and $E(k) \sim k^{-5/3}$. The Wiener-Khinchin theorem, frequently applied in signal processing, formalizes this connection through the identity $2/3 + 1 = 5/3$. However, in BDT, no clear relationship between these scaling laws emerges for the statistics of the original velocity fields. This study investigates the underlying factors causing this discrepancy and aims to reconstruct the scaling relationship between the energy spectrum and $S_2(\ell)$.

The bulk flow of BDT exhibits significant anisotropy, as evidenced by the large-scale flow structures observed in numerical simulations (Fig. 1), even at sufficiently high Reynolds numbers. To mitigate the influence of large flow structures on the structure functions, it is necessary to filter out the low-wavenumber components of the energy spectrum. To do so, we reconstruct the velocity field by retaining only the high wavenumber helical modes within the two red triangle symbols shown in Fig. 2 (Methods). Fig. 3 illustrates the compensated plot of $S_2(\ell)/\ell$ and $-D_3(\ell)/\ell$ for the filtered field, where $D_3(\ell) \equiv \langle (\delta v_\ell)^3 \rangle$ representing the third-order moments of $\delta v_\ell$—without taking the absolute value. In all BDT cases examined, the scaling behavior of $D_3(\ell)$ approximates $D_3(\ell) \sim \ell$. For SHF and the high Reynolds number cases of SVK and SDF, $D_3(\ell)$ closely follows the 4/5 law, expressed as $D_3(\ell) \approx -4/5 \langle \epsilon \rangle \ell$, where $\langle \epsilon \rangle$ is the averaged energy dissipation rate. This law represents the only exact result derived from Navier-Stokes equations for homogeneous turbulence[2]. Notably, an effective scaling of $S_2(\ell) \sim \ell$ is conspicuous within the identical scaling range of $D_3(\ell)$. Fig. S2 clearly demonstrates the linear relationship between $S_2(\ell)$ and $D_3(\ell)$ in the inertial range for each BDT case.

At lower Reynolds numbers, the scaling range is not readily discernible in the log-log plot of $S_2(\ell)$ versus $\ell$, even for the filtered fields. To enhance the scaling range and facilitate the identification of scaling exponents, we apply the extended self-similarity (ESS)[8] (Fig. S3), which plot $S_p(\ell)$ versus $S_2(\ell)$, where $S_p(\ell) \equiv \langle |\delta v_\ell|^p \rangle$ denotes the $p$-order statistical moments of the absolute value of velocity increments. The scaling range in ESS corresponds to the shaded region depicted in both Fig. 2 and Fig. 3, encompassing the platform area in the compensated plot (Fig. 3) and the power-law range in the energy spectrum (Fig. 2). This alignment provides compelling evidence that the scaling exponent of $S_2(\ell)$ is 1, which is consistent with the $k^{-2}$ power-law of the energy spectrum in the inertial range. Furthermore, we observe that in most cases, $S_2(\ell)$ exhibits a noticeable scaling range in both the original (Fig. 2) and the filtered fields (Fig. 3). However, the scaling exponents in these two fields differ significantly. This discrepancy suggests that the structure functions in the original anisotropic flows do not correctly capture the small-scale statistical behavior in the inertial range. By applying appropriate filtering techniques, we restore the relationship between the scaling exponents of $S_2(\ell)$ and the energy spectrum.

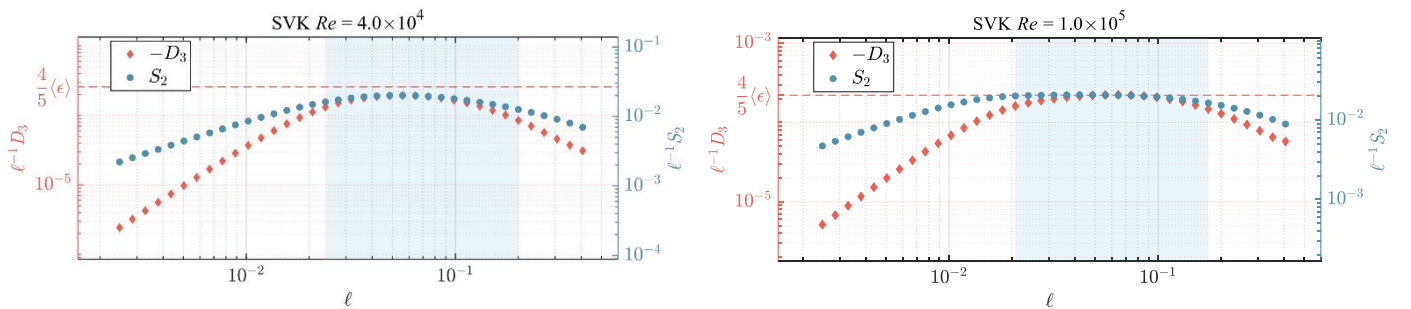

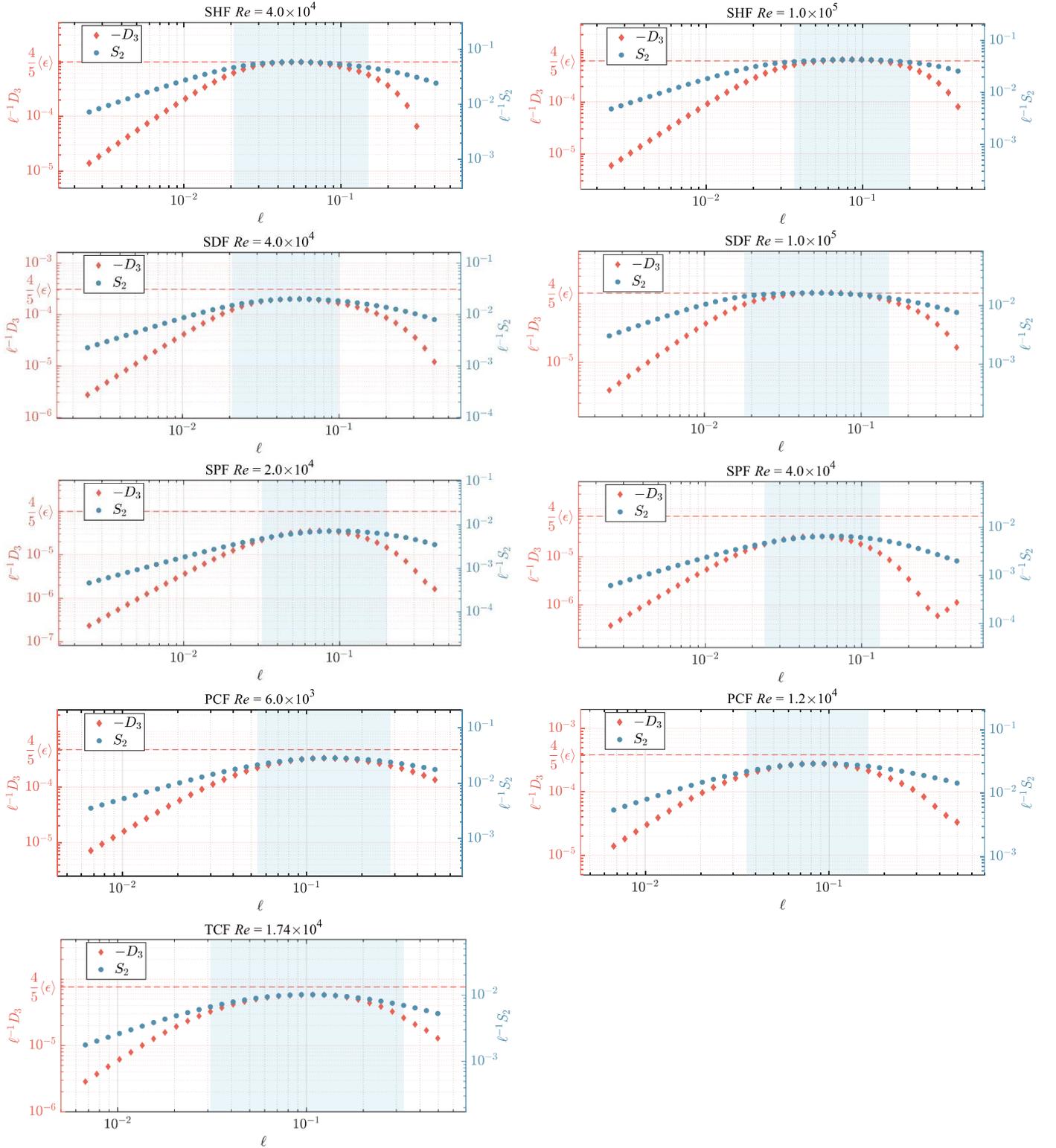

**Fig. 3. Compensated structure functions of $S_2(\ell) \equiv \langle |\delta v_\ell|^2 \rangle$ (solid blue circles) and $D_3(\ell) \equiv \langle (\delta v_\ell)^3 \rangle$ (solid red squares) for the filtered velocity field.** In all panels, the red dashed line represents $4/5\langle \epsilon \rangle$ while the blue shaded region corresponds to the scaling range derived from the ESS, where $\langle \epsilon \rangle$ denotes the averaged energy dissipation rate for filtered field. $S_2(\ell)$ and $D_3(\ell)$ are compensated by $\ell$ and plotted on separate axes. The vertical axis for $D_3/\ell$ is on the left, while the vertical axis for $S_2/\ell$ is on the right. The horizontal axis is shared between both plots. For convenience, the vertical axis ranges for each function are adjusted so that $D_3/\ell$ and $S_2/\ell$ align along the same horizontal line, facilitating easier comparison of the platform area.

Anisotropy, induced by large-scale flow structures and low-wavenumber components, likely drives the observed inconsistency between the scaling exponents of energy spectra and the second-order structure functions. This anisotropy generates variability in the scaling behavior of the $S_2(\ell)$, disrupting the correspondence between structure function and energy spectrum exponents, even at high Reynolds numbers. This hypothesis is further supported by comparing the degree of isotropy before and after filtering.

We define a local ratio of kinetic energy components to total kinetic energy within a sphere as

$$I_i(\boldsymbol{x}, t) = \frac{\langle u_i^2(\boldsymbol{y}, t)\rangle_{\boldsymbol{y}\in\text{Ball}(\boldsymbol{x},\ell_0)}}{\langle u_1^2(\boldsymbol{y},t) + u_2^2(\boldsymbol{y},t) + u_3^2(\boldsymbol{y},t)\rangle_{\boldsymbol{y}\in\text{Ball}(\boldsymbol{x},\ell_0)}}, \qquad i = 1, 2, 3,$$

where $\langle A\rangle_{\boldsymbol{y}\in\text{Ball}(\boldsymbol{x},\ell_0)}$ represents the average of variable $A$ over a spherical region centered at $\boldsymbol{x}$ with diameter $\ell_0$, the largest length scale of the inertial range. On this basis, a scalar index $s(\boldsymbol{x}, t)$ is defined as

$$s(\boldsymbol{x}, t) = \frac{(I_1 - I_2)^2 + (I_2 - I_3)^2 + (I_3 - I_1)^2}{3},$$

which quantitatively assess the isotropy of the spatial distribution of local kinetic energy. The value of $s(\boldsymbol{x}, t)$ ranges from 0 and 2/3, with a distribution centered around 0 indicating local isotropy. Indeed, the distribution of $s(\boldsymbol{x}, t)$ for HIT ($Re_\lambda = 315$, see Methods) as shown in Fig.4 is characterized by a sharp peak near 0, with a short tail (grey solid lines), suggesting that the flow field is approximately isotropic.

Using HIT as a benchmark, we evaluate the isotropy of the flow in BDT. Fig. 4 shows the PDFs of $s(\boldsymbol{x}, t)$ at various Reynolds numbers, both before and after filtering, for each flow condition. Anisotropy in BDT is evident before filtering, with the exception of SVK. After filtering, however, the degree of local isotropy improves markedly, resulting in a narrow peak near 0, even surpassing the isotropy observed in HIT. These results confirm the effectiveness of the HWD-based filtering method in removing anisotropic components from the original turbulent flow.

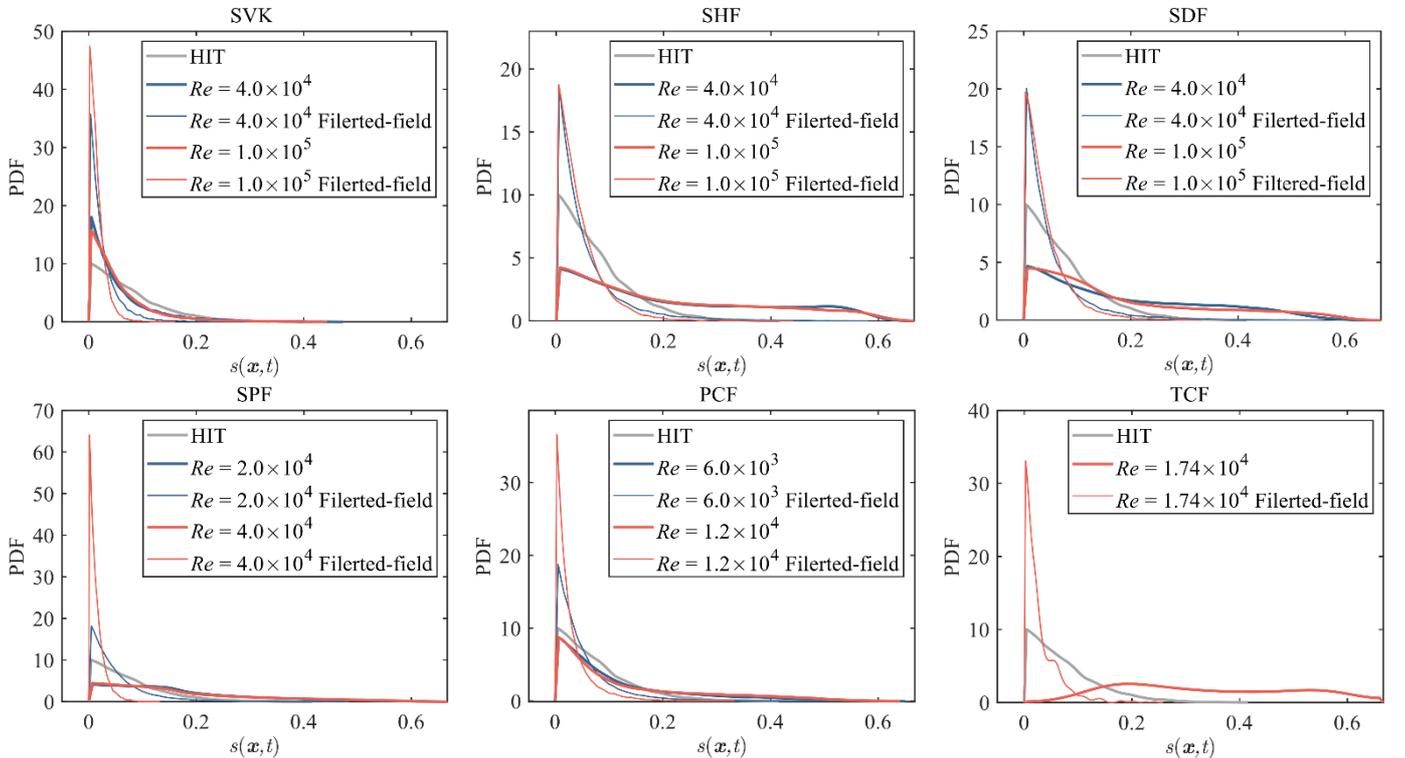

**Fig. 4. Comparison of local isotropy before (thick solid lines) and after (thin solid lines) filtering.** The PDFs of the index $s(\boldsymbol{x}, t)$ are shown for various BDTs at different Reynolds numbers. All panels include HIT (grey solid line) for comparison. The red lines correspond to cases with high Reynolds numbers, while the blue lines represent cases with low Reynolds numbers.

In the Taylor-Couette apparatus, Ezeta *et al.*[21] performed particle image velocimetry experiments to measure

$S_2(\ell)$ of azimuthal velocity in bulk flow across a range of Taylor numbers $4.0 \times 10^8 < Ta < 9.0 \times 10^{10}$, corresponding to the ultimate regime of TCF. Their results indicated that $S_2(\ell)$ followed a 2/3 power law in the compensated plot (Fig. 5A), similar to our findings for the non-filtering field of DNS. Interestingly, another scaling range is apparent on the left side of the platform region in Fig. 5A. Re-plotting the compensated plot of $S_2(\ell)/\ell$ (Fig. 5B) shows that the scaling exponent is closer to 1, consistent with our post-filtering results based on DNS for TCF at $Ta = 4.62 \times 10^8$ ($Re_i = 1.74 \times 10^4$). Minor deviations to 1 are observed for some instances ($Ta = 1.6 \times 10^9, 3.6 \times 10^9, 6.4 \times 10^9$), which are likely due to the influence of low-wavenumber components within the flow. Similar phenomena are observed in the high Reynolds number case of SVK, SHF and SDF, where the scaling exponent of $S_2(\ell)$ in the non-filtered field deviates slightly from 1.

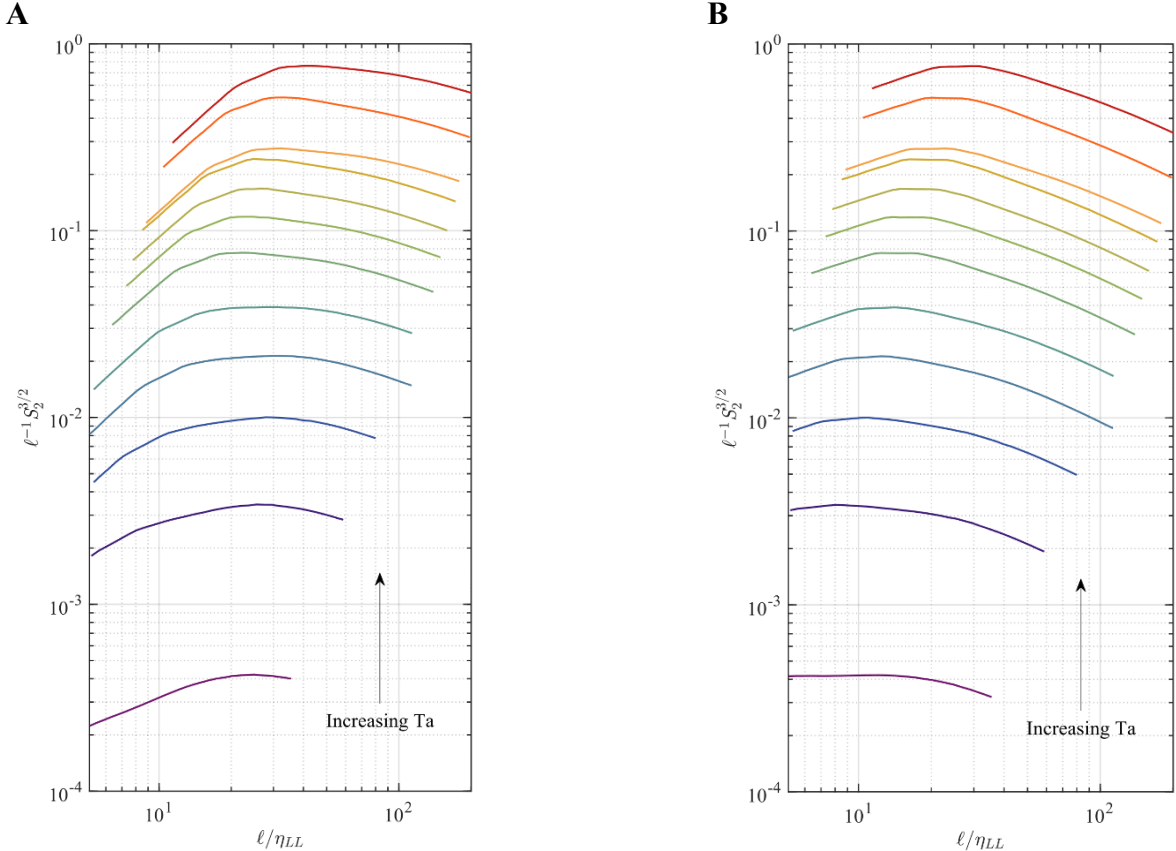

**Fig. 5. Compensated second-order structure functions of azimuthal velocity component for TCF experiments in various Ta numbers**[21]. Each solid line, from bottom to top, represents the following Taylor numbers ($Ta$): $4.0 \times 10^8$, $1.6 \times 10^9$, $3.6 \times 10^9$, $6.4 \times 10^9$, $1.0 \times 10^{10}$, $1.4 \times 10^{10}$, $2.0 \times 10^{10}$, $2.6 \times 10^{10}$, $3.2 \times 10^{10}$, $4.0 \times 10^{10}$, $5.7 \times 10^{10}$, $9.0 \times 10^{10}$. **A,** Longitudinal, bulk-averaged structure functions ($S_2(\ell) \equiv \langle |u_\theta(r, \theta + \ell/r, t) - u_\theta(r, \theta, t)|^2 \rangle$) in the bulk region are compensated as $\ell^{-1} S_2^{3/2}$ and plotted against dimensionless length $\ell/\eta_{LL}$ (where $\eta_{LL}$ is the Kolmogorov length scale). This data is adapted from Ezeta *et al.*[21]. **B,** Replot of $S_2(\ell)$ compensated as $\ell^{-1} S_2$ versus $\ell/\eta_{LL}$. For easier comparison, the curves are vertically shifted.

DNS and experimental results demonstrate that the second-order velocity structure functions of BDT follow a distinct power law compared to HIT. Nevertheless, $D_3(\ell)$ continues to obey the 4/5 law. These observations imply that an unknown physical quantity, other than the dissipation rate, may dominate the energy cascade in the inertial range of BDT.

## Scaling laws of high-order structure functions

In this section, we present a phenomenological prediction for the scaling exponents of $S_p$ based on the Kolmogorov similarity hypothesis[3,27] and a multiscale model for the PDFs of $\delta v_\ell$ within the scaling range.

This PDF model is used to predict the scaling laws of $D_p \equiv \langle (\delta v_\ell)^p \rangle$, the $p$th-order moment of $\delta v_\ell$ for odd integers $p$. Since $S_2(\ell)$ and $D_3(\ell)$ both scale linearly with $\ell$ in the inertial range, the ratio $-D_3/S_2$ can be interpreted as a characteristic speed $U$, which remains invariant across scales within this range (Fig. S3). In BDT, the bulk flow is sustained by energy injected through the active boundary layer. At large Reynolds numbers, this energy injection can be quantitatively assessed via the wall-normal velocity at the interface between the bulk region and the boundary layer. The average of the absolute values of this velocity over the bulk-BL (boundary layer) interface defines a "projectile speed", reflecting both the strength of fluctuations at the interface and the power driving the bulk flow. We show that the characteristic velocity extracted from the inertial range is directly linked to the projectile speed at the interface (Table S1).

According to Kolmogorov's refined similarity hypothesis[27], the coarse-grained dissipation rate $\epsilon_\ell$ and velocity increments $\delta v_\ell$ are related by

$$\epsilon_\ell \sim (\delta v_\ell)^2 / \tilde{t}, \tag{1}$$

where $\tilde{t}$ denotes the typical time scale. Generally, $\tilde{t}$ is represented by

$$\tilde{t} \sim \ell / |\delta v_\ell| . \tag{2}$$

Following Eqs. (1) and (2), we can derive the magnitude of $|\delta v_\ell|$ as

$$|\delta v_\ell| \sim \epsilon_\ell^{1/3} \ell^{1/3}. \tag{3}$$

For BDT, we propose an alternative timescale

$$\tilde{t} \sim \ell / U. \tag{4}$$

Substituting Eq. (4) into Eq. (1) yields

$$|\delta v_\ell| \sim (\epsilon_\ell / U)^{1/2} \ell^{1/2}. \tag{5}$$

Such analysis produces $S_2(\ell) \sim \langle \epsilon_\ell / U \rangle \ell$ immediately. Furthermore, the scaling behavior of the $p$th-order statistical moment of $|\delta v_\ell|$ and $\epsilon_\ell$ is described by

$$\langle |\delta v_\ell|^p \rangle \sim \ell^{\zeta_p}, \qquad \langle \epsilon_\ell^p \rangle \sim \ell^{\tau_p}. \tag{6}$$

By integrating Eqs. (3), (6) and Eqs. (5), (6), respectively, the relationship between $\zeta_p$ and $\tau_p$ is expressed as

$$\zeta_p = p/3 + \tau_{p/3}, \qquad \text{for HIT}, \tag{7}$$

and

$$\zeta_p = p/2 + \tau_{p/2}, \qquad \text{for BDT}. \tag{8}$$

A phenomenological ansatz for the structure functions, based on the concept of hierarchical structures[28], leads to the expression for $\tau_p$ as

$$\tau_p = -2p/3 + 2[1 - (2/3)^p], \tag{9}$$

where the coefficients are derived from reasonable physical arguments rather than the adjustable parameters. Substituting this expression for $\tau_p$ (Eq. (9)) into the equation for BDT scaling (Eq. (8)), we obtain the scaling law for $S_p(\ell)$

$$\zeta_{p,BDT} = p/6 + 2[1 - (2/3)^{p/2}]. \tag{10}$$

For comparison, the scaling law for $S_p(\ell)$ in HIT, known as the She-Leveque scaling[28], is given by

$$\zeta_{p,HIT} = p/9 + 2[1 - (2/3)^{p/3}]. \tag{11}$$

This scaling law is derived from Eqs. 7 and 9. Both BDT and HIT exhibit the same coarse-grained energy dissipation rate scaling laws, suggesting that the most intense dissipation events in BDT are one-dimensional vortex filaments, similar to those in HIT[28,29]. These intense events, however, are independent of the characteristic speed $U$ in BDT. Despite this similarity, the relationship between $\delta v_\ell$ and $\epsilon_\ell$ in BDT and HIT differs fundamentally, as shown in Eqs. (7) and (8). This comparison highlights the fundamental differences in the energy dissipation mechanisms between BDT and HIT, which has significant implications for understanding the small-scale dynamics of turbulence at high Reynolds numbers.

Fig. 6 compares the scaling exponents $\zeta_p$ and $\tau_p$ from DNS with the phenomenological predictions (Eqs. (9) and (10)). In the spherical-wall cases, the value of $\tau_p$ agrees closely with the theoretical prediction (Eq. (9)), as shown in Fig. 6A. For all the BDT cases studied, the scaling exponents of $S_p(\ell)$ (for $p = 1, \ldots, 9$) align excellently with Eq. (10) (Fig. 6B), with a maximum relative error no more than 4% (Table S2). Fig. 6B clearly demonstrates that $\zeta_{p,BDT}$ diverges significantly from $\zeta_{p,HIT}$ as the $p$ increases, a trend consistent with previous studies on TCF[10]. The exponents in these studies are recalculated as $\zeta_p = \zeta_p/\zeta_2$, showing excellent agreement with Eq. (10). Therefore, the phenomenological model accurately describes the DNS and experimental data across various BDT types. Notably, the model applies only to $S_p(\ell)$ as the statistical moment of the absolute value of $\delta v_\ell$, and does not account for the observed non-negligible difference between the scaling exponents of $S_3(\ell)$ and $D_3(\ell)$—a distinction not observed in HIT.

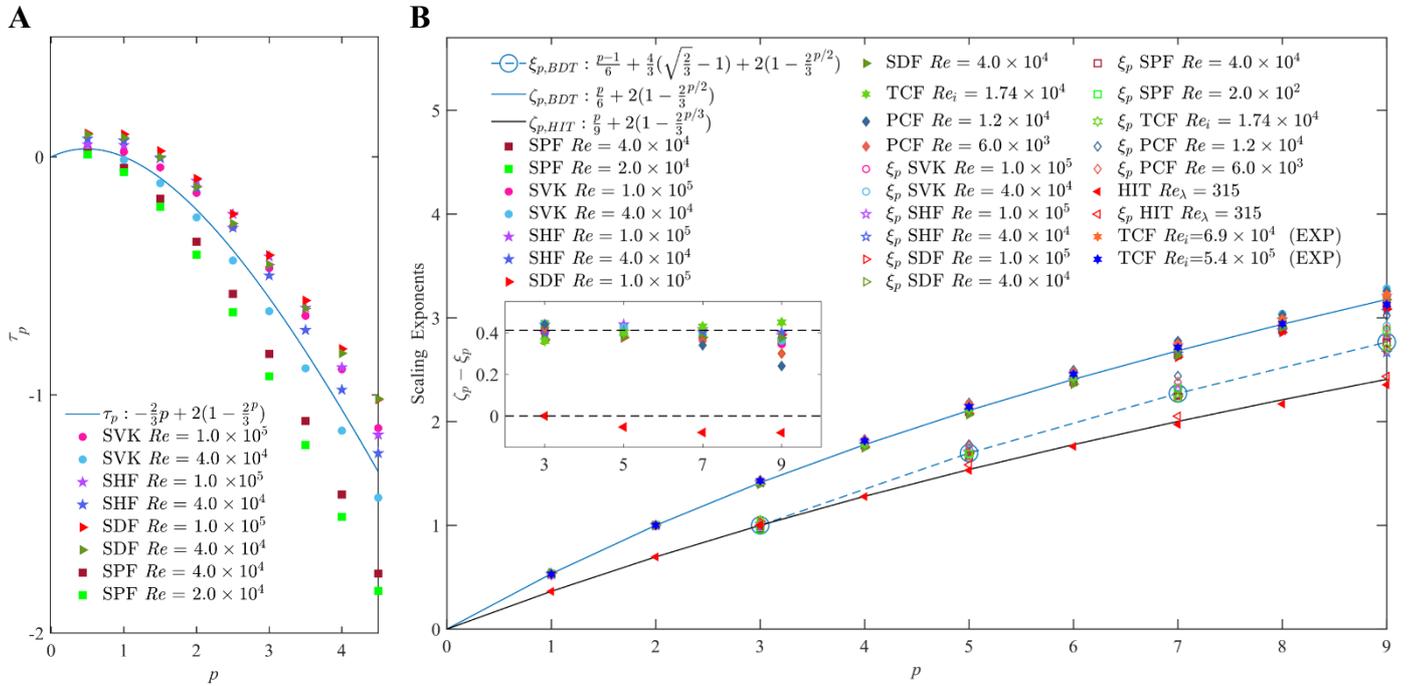

**Fig. 6. Scaling exponents of the coarse-grained dissipation rate and the longitudinal velocity structure functions. A,** Scaling exponents $\tau_p$ of $p$th-order moments of the coarse-grained dissipation rate $\epsilon_l$ as a function of $p$. The phenomenological prediction (blue solid line) and DNS (solid symbols) for spherical domain with various boundary conditions are illustrated. **B,** Scaling exponents $\zeta_p$ of $p$th-order moments of the absolute velocity increments $|\delta v_\ell|$ as a function of $p$, and exponents $\xi_p$ of $p$th-order moments of the velocity increments $\delta v_\ell$ as a function of odd $p$. For BDT cases, the phenomenological prediction (blue lines; solid line for $\zeta_p$ and dash line with open circles for $\xi_p$) and DNS results (symbols; solid symbols for $\zeta_p$ and open symbols for $\xi_p$) are presented. Results from the TCF experiments[10] (solid blue and orange hexagons) are analyzed using time series data. The HIT case is also shown for comparison (solid red triangle symbols for $\zeta_p$, open red triangle symbols for $\xi_p$, and solid black line for theoretical model). The inset highlights the differences between $\zeta_p$ and $\xi_p$ as a function of odd $p$ for BDT and HIT, emphasizing the discrepancies between DNS results (symbols) and phenomenological predictions (black dashed lines).

Next, we investigate the differences in scaling exponents of $S_p(\ell)$ and $D_p(\ell)$ for odd values of $p$, aiming to understand the behavior of the PDFs of $\delta v_\ell$ within the scaling range. Under the approximation of local homogeneity and isotropy, the first moments of $\delta v_\ell$ vanish, and for large $\ell$, the distribution of $\delta v_\ell$ is expected to approximate a Gaussian distribution. However, experimental data show that higher odd-order moments of $\delta v_\ell$ do not vanish, suggesting that the distribution deviates from a purely Gaussian form. To better capture the characteristics of the PDFs of $\delta v_\ell$, we use a Gaussian mixture model (GMM), implemented as a linear superposition of multiple Gaussians. It was found that two Gaussians are sufficient to accurately describe the PDF of $\delta v_\ell$ at the largest scale $\ell_0$ within the inertial range.

We propose a model for the PDF of velocity increments $\delta v_\ell$ in the scaling range, defined as

$$P(\delta v_\ell) = \sum_{i=1}^{2} \sum_{k=0}^{\infty} e^{-\lambda} \frac{\lambda^k}{k! \, (\ell/\ell_0)^\gamma \beta^k} \frac{m_i}{\sqrt{2\pi}\sigma_i} \exp\left[-\frac{(\delta v_\ell - \mu_i(\ell/\ell_0)^\eta \beta^k)^2}{2\sigma_i^2 (\ell/\ell_0)^{2\gamma}\beta^{2k}}\right], \qquad \lambda = -C\ln(\ell/\ell_0). \quad (12)$$

In this model, $C, \gamma, \beta$ and $\eta$ are constants associated with scaling exponents (see the following Eqs. 16 to 17), while $m_i, \mu_i$, and $\sigma_i$ depend on the distribution of $\delta v_{\ell_0}$ in the flow. It is specified that $\ell_0 \geq \ell$. When $\ell = \ell_0$, the PDF of $\delta v_{\ell_0}$ is given by

$$P(\delta v_{\ell_0}) = \sum_{i=1}^{2} \frac{m_i}{\sqrt{2\pi}\sigma_i} \exp\left[-\frac{(\delta v_{\ell_0} - \mu_i)^2}{2\sigma_i^2}\right], \quad (13)$$

where each Gaussian is characterized by three parameters: a mean $\mu_i$, a variance $\sigma_i$ and a mixing weight $m_i$. The model corresponds to a dual log-Poisson cascade[30,31], enabling mapping of any inertial-range velocity increment distribution $P(\delta v_\ell)$ (for $\ell < \ell_0$) from the parent distribution $P(\delta v_{\ell_0})$. Although the number of Gaussians appears to be infinite for $\ell < \ell_0$, the term $\lambda^k/k!$ decays rapidly as $k$ increases, indicating that a finite number of Gaussian functions can effectively describe the PDFs of $\delta v_\ell$ in practice.

The $p$th-order moments of $|\delta v_\ell|$ and the odd $p$th-order moments of $\delta v_\ell$ can be derived from Eq. (12), yielding

$$S_p(\ell) = \sqrt{2^p/\pi} \, \Gamma\left(\frac{p+1}{2}\right) (\ell/\ell_0)^{p\gamma + C(1-\beta^p)} \sum_{i=1}^{2} m_i \sigma_i^p \,, \quad (14)$$

$$D_p(\ell) = \sqrt{2^{p+1}/\pi} \, \Gamma\left(\frac{p+2}{2}\right) (\ell/\ell_0)^{(p-1)\gamma + \eta + C(1-\beta^p)} \sum_{i=1}^{2} m_i \mu_i \sigma_i^{p-1} \,. \quad (15)$$

The GMM model captures the characteristics of the distribution of $\delta v_\ell$, where the first moments are zero and the higher moments are negative. These equations correspond to the scaling law, $S_p(\ell) \sim \ell^{\zeta_p}$ and $D_p(\ell) \sim \ell^{\xi_p}$, respectively, where

$$\zeta_p = p\gamma + C(1-\beta^p), \quad (16)$$

$$\xi_p = (p-1)\gamma + \eta + C(1-\beta^p) = \zeta_p + (\eta - \gamma) \text{ for odd } p. \quad (17)$$

From the 4/5 law, $\xi_3 = 1$, so it follows that

$$\eta = 1 - 2\gamma - C(1-\beta^3).$$

From Eqs. 10 to 11, we obtain

$$C = 2, \qquad \gamma = \eta = 1/9, \qquad \beta = (2/3)^{1/3}, \text{ for HIT}, \quad (18)$$

$$C = 2, \qquad \gamma = 1/6, \qquad \eta = 4(\sqrt{2/3}-1)/3, \qquad \beta = \sqrt{2/3}, \qquad \text{for BDT}. \quad (19)$$

These parameters lead to the explicit formulae for the full set of scaling exponents of velocity structure function of the BDT, as appeared in Fig. 6B.

From Eqs. 16 to 18, it is evident that the scaling exponents of $S_p(\ell)$ and $D_p(\ell)$ are identical in HIT. This is consistent with the experimental and numerical observation that $\zeta_3 = \xi_3$, suggesting that the scaling exponents $\zeta_p$ and $\xi_p$ will remain identical for larger odd $p$ (Fig. 6B). In contrast, the scaling behavior of the mean and variance of each Gaussian in Eq. (12) with respect to $\ell$ differs in BDT. Specifically, a constant difference between $\zeta_p$ and $\xi_p$ for odd $p$ (approximately $\gamma - \eta \approx 0.41$) is observed, reflecting distinct scaling behavior in BDT compared to HIT. The scaling exponents of $D_p(\ell)$ for the cases are presented in Fig. 6B. The inset, reveals a notable deviation of $\zeta_p - \xi_p$ from 0.41 at high orders (here $p = 9$) in the PCF case. This deviation can likely be attributed to the insufficient Reynolds number or other unknown reasons in the analyzed cases.

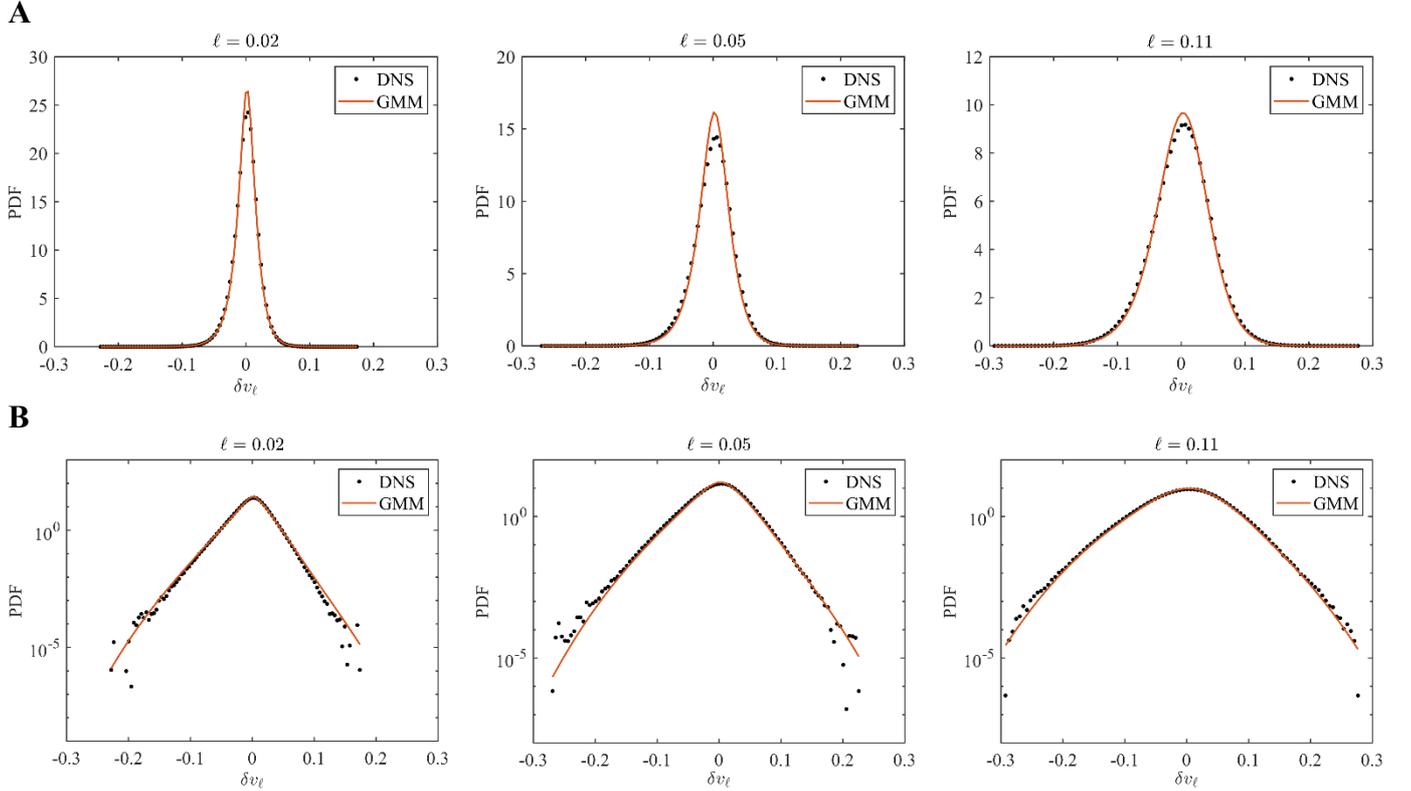

**Fig. 7. Comparison of the PDFs of velocity increments in the inertial range: DNS v.s. model (GMM).** The PDFs of velocity increments $\delta v_\ell$ for the SVK case ($Re = 1.0 \times 10^5$) are compared between DNS results (symbols) and the model (solid lines) across various length scale $\ell$. The values of $\ell$ increases from left to right, with specific values of 0.02, 0.05 and 0.11. **A,** Linear plot. **B,** Semi-logarithmi plot.

For the SVK case at $Re = 1.0 \times 10^5$, we apply the iterative Expectation-Maximization (EM) algorithm to determine $m_i, \mu_i,$ and $\sigma_i$ at $\ell_0 = 0.15$ (Fig. S5 and Table S3). The PDFs of $\delta v_\ell$ at various $\ell$ show excellent agreement with the GMM given by Eq. (12) (Fig. 7). In the linear plot (Fig. 7A), minor discrepancies between the GMM and DNS near $\delta v_\ell = 0$ are observed. However, these differences are too small to influence the scaling exponents, even for low-order structure functions. The tail behavior is also well captured by Eq. (12) in the semi-logarithmic plot (Fig. 7B), further validating the phenomenological prediction (Eqs. (10) and (17)) for high-order structure functions. These results demonstrate that the GMM effectively models the PDF of $\delta v_\ell$ in the inertial range for BDT, and may also be applicable to other flow types. By allowing for non-zero higher odd-order moments, the GMM resolves the discrepancy between Gaussian assumptions and experimental observations. This provides a more accurate description of the scaling behavior of $\delta v_\ell$, especially for odd-order moments critical to turbulence dynamics.

## Discussion and Conclusion

In recent years, turbulence research has increasingly focused on complex fluids — including quantum[32,33], elastic[34,35], viscoelastic[36–38], and even nematic turbulence[39] — alongside the broader category of multiphase flows. However, classical turbulence in Newtonian fluids remains a fundamental challenge in physics. For Newtonian turbulence, energy injection arises from external body forces (e.g., buoyancy, electromagnetic forces), internal pressure gradients (gravity and other conservative forces can also be included wherein), or boundary motions (e.g., flying objects, turbomachinery). Kinetic energy injected at large scales cascades to smaller scales, ultimately dissipating as heat. This energy cascade underpins Kolmogorov's phenomenology, which remains the most successful turbulence theory despite its assumption of local isotropy (excluding anisotropic flows like buoyancy-driven turbulence). While the 2/3 and −5/3 scaling laws hold for a majority of turbulent flows, the boundaries of their validity under the prerequisites of this theory remain unclear. Aiming at this question, we concentrate mainly on the scaling laws of velocity structure functions and the energy spectra in the large for some novel Newtonian turbulence, along with a few canonical turbulent flows not fully explored previously. Specifically, the small-scale statistics in the bulk flows of some typical BDT are investigated. To ensure the greatest possible robustness of statistical average, the global spatiotemporal statistics both for the structure functions and energy spectrum is employed. Such substantial effort is unusual but has yielded promising results.

Our findings suggest that the bulk flows of BDT do not follow the traditional 2/3 and −5/3 scaling laws, but follow an alternative 1 and -2 scaling laws, revealing non-Kolmogorov universal scaling laws in the inertial range. Moreover, the higher-order velocity structure functions obey a non-linear scaling law similar to those found in numerous turbulent flows, but the parameters in scaling laws are evidently distinct. In particular, the odd-order structure functions with and without absolute value have different scaling exponents. To understand these peculiar behaviors, we proposed a phenomenological theory based on the energy dissipation rate and a characteristic speed, alongside a model for the PDF of the velocity increments $\delta v_\ell$. It is believed that the projectile motion at the interface between the bulk and boundary layer play a pivotal role for the small-scale statistics in the bulks of BDT. A comprehensive study on this problem as well as other case study at much higher Reynolds numbers will be carried out in the future.

Notably, for BDT, no stirring body force is exerted, and small-scale isotropy can be restored through high-pass filtering. Thus, there seems no apparent reason to reject the prevailing 2/3 and −5/3 scaling laws in BDT. However, our findings challenge this presumption and provide a new perspective on the small-scale statistical behavior of BDT. Therefore, the present work may lay the groundwork for extending turbulence study beyond the original Kolmogorov 1941 theory. Moreover, these new understandings could have broader implications for industrial flows and some geophysical and astrophysical turbulence, where boundary driving is critical.

## Acknowledgements

This work was supported by the National Natural Science Foundation of China (NSFC) under Grant Nos.11672005, 11521091, 11221062. The data for this study were obtained using the Tianhe-2 supercomputer at the National Supercomputer Center in Guangzhou, China.

## Author contributions

Y.Z. and W.S. conceived the ideas. Y.Z., Z.L. and J.L. performed the numerical simulations. Y.Z. analyzed the data. Y.Z. and W.S. wrote the paper. W.S. supervised the project. All authors proofread the paper.

## Competing interests

The authors declare no competing financial interests.

## Methods

### Numerical methods

DNSs of the incompressible Navier-Stokes equations are employed to obtain high-fidelity datasets of flow fields.

*Spherical domains.* The HWD-based pseudo-spectral method devised by Liao & Su[40] is employed to solve the governing equations in spherical domain cases. The 3rd-order Runge-Kutta algorithm is applied for temporal advancement, while the fast algorithm for spherical harmonic transform and Gaussian quadrature are employed to compute the nonlinear term in the NS equations. The domain radius $a$ is set to 1. Four distinct cases (SVK, SHF, SDF, and SPF) are simulated, each with varying steady velocity distributions prescribed on the boundary. For each case, simulations are conducted at two different Reynolds numbers (Table S4). The Reynolds number is defined as $Re = U_c a/\nu$, where $U_c$ is the typical velocity at the boundary (set to 1) and $\nu$ is the kinematic viscosity. The grid resolution for each case is $N_\phi \times N_\theta \times N_r = 512 \times 512 \times 432$.

*Channel domains.* The code developed by Yang[41] is used to solve the PCF cases in a Cartesian coordinate, where the streamwise, wall-normal, and spanwise directions are denoted by $x$, $y$, and $z$, respectively. Spatial discretization is achieved using a Fourier-Chebyshev expansion. The explicit Adams-Bashforth scheme is used for the time integration of the convective terms, while the implicit Crank-Nicolson scheme is applied for the diffusive terms. The flow is driven by two parallel planes moving in opposite tangential directions at a constant velocity $U_w$. The DNS are conducted within a finite rectangular domain with periodic boundary conditions in the $x$ and $z$ directions, with domain dimensions $L_x \times 2h \times L_z = 32 \times 1 \times 8$. The grid resolutions are $1280 \times 240 \times 768$ and $2560 \times 384 \times 1563$ for $Re = 6.0 \times 10^3$ and $Re = 1.2 \times 10^4$, respectively, where $Re = 2U_w h/\nu$ with $U_w = 1$.

*Cylindrical domains.* DNS of TCF in cylindrical coordinates is performed using the nsCouette code developed by Lopez[42]. In this case, the radial, axial, and azimuthal directions are denoted by $r$, $z$ and $\theta$, respectively. Simulations are conducted at a Taylor number $Ta = 4.62 \times 10^8$ (with inner and outer Reynolds numbers $Re_i = 1.74 \times 10^4$, $Re_o = 0$). The predictor-corrector method is used for time advancement, while Fourier-Galerkin discretization and high-order finite differences are employed in the radial and wall-parallel directions, respectively. The Taylor number is calculated as $Ta = [(1 + \eta)^4/(64\eta^2)]d^2(r_i + r_o)^2(\omega_i - \omega_o)^2\nu^{-2}$, with $r_{i,o}$ the radius of the inner and outer cylinders, $\eta = r_i/r_o = 0.714$ the ratio between the inner and outer cylinder radius, $d = r_o - r_i = 1$ the gap width and $\omega_{i,o}$ the angular velocities of the inner and outer cylinders. $Re_{i,o} = U_{i,o}d/\nu$ denote the inner and outer Reynolds numbers, where $U_{i,o}$ are the velocities of the inner and outer cylinders. The system is characterized by a rotational symmetry of order $n_{sym} = 6$, and an aspect ratio of $\Gamma = H/d = 2\pi/3$, where $H$ represents the domain height. The simulation resolution is $N_\theta \times N_r \times N_z = 382 \times 640 \times 577$.

*HIT.* The DNS raw data for HIT used in this study were downloaded from a database developed by Jiménez's group[43].

<u>Energy spectrum for arbitrary domains</u>

The three-dimensional incompressible velocity field $\boldsymbol{u}(\boldsymbol{x}, t)$ in properly smooth bounded domain can be expressed in terms of complete and orthogonal helical-wave vector bases $\boldsymbol{B}_k$, which serves as the eigenfunctions of the curl operator[44]. This leads to the relationship

$$\begin{cases} \nabla \times \boldsymbol{B}_k = \lambda_k \boldsymbol{B}_k, & \text{in } \Omega, \\ \boldsymbol{n} \cdot \boldsymbol{B}_k = 0, & \text{on } \partial\Omega, \end{cases} \tag{20}$$

where $\lambda_k$ represents the eigenvalue, $\Omega$ denotes the flow domain, and $\boldsymbol{n}$ is the wall-normal vector of the domain boundary. For general boundary conditions with $\boldsymbol{u} \cdot \boldsymbol{n} \neq 0$, the velocity field $\boldsymbol{u}$ can be uniquely represented as a superposition of $\boldsymbol{B}_k$ and a potential field $\varphi$, such that

$$\boldsymbol{u} = \nabla\varphi + \sum_k c_k \, \boldsymbol{B}_k,$$

where $c_k$ are the expansion coefficients, and $\boldsymbol{u} \cdot \boldsymbol{n}|_{\partial\Omega} = \partial\varphi/\partial n$ with $\nabla^2\varphi = 0$. Since $\boldsymbol{B}_k$ form an orthogonal set, the coefficients $c_k$ are given by

$$c_k = \int_\Omega \boldsymbol{u} \cdot \boldsymbol{B}_k^* \, dV,$$

where the asterisk denotes the complex conjugate when complex numbers are introduced into $\boldsymbol{B}_k^*$. The global energy spectrum $E(k)$ is defined in terms of $c_k$ by the expression

$$E(k) = \frac{1}{2V} \sum c_k c_k^*, \tag{21}$$

where $V$ represents the domain's volume. This energy spectrum generalizes the conventional Fourier spectrum and is applicable to any bounded three-dimensional domain[45]. By applying the curl operation to Eq. (20), we derive the vector Helmholtz equation

$$(\nabla^2 + \lambda_k^2)\boldsymbol{B}_k = 0.$$

We use the formular provided by Morse and Feshbach[46] to express $\boldsymbol{B}_k$ as

$$\boldsymbol{B}_k = \nabla \times (\boldsymbol{e}w\psi_k) + \frac{1}{\lambda_k}\nabla \times \nabla \times (\boldsymbol{e}w\psi_k),$$

where $\boldsymbol{e}$ is the unit vector and is identical to $\boldsymbol{e}_i$, $\boldsymbol{e}_r$ and $\boldsymbol{e}_z$ in Cartesian, spherical, and cylindrical coordinates, respectively. The corresponding scalar $w$ takes value $1$, $r$, and $1$ in these coordinate systems, respectively. The function $\psi_k$ satisfies the Helmholtz equation

$$(\nabla^2 + \lambda_k^2)\psi_k = 0. \tag{22}$$

Solving Eq. (22) in various geometric domains yields the specific expression for the helical-wave bases.

*Spherical domain solution.* In a spherical domain with radius $a$, the helical-wave $\boldsymbol{B}_k$ is expressed in spherical coordinates as[40]

$$\begin{pmatrix} B_{qlm}^r \\ B_{qlm}^\theta \\ B_{qlm}^\phi \end{pmatrix} = \frac{1}{|\lambda_{ql}j_{l+1}(|\lambda_{ql}|a)|\sqrt{l(l+1)a^3}} \begin{pmatrix} \dfrac{l(l+1)}{r}j_l(|\lambda_{ql}|r)Y_{lm}(\theta,\phi) \\ \lambda_{ql}j_l(|\lambda_{ql}|r)\dfrac{imY_{lm}(\theta,\phi)}{\sin\theta} + \dfrac{1}{r}\dfrac{\partial}{\partial r}\left(rj_l(|\lambda_{ql}|r)\right)\dfrac{\partial}{\partial\theta}Y_{lm}(\theta,\phi) \\ \lambda_{ql}j_l(|\lambda_{ql}|r)\dfrac{\partial}{\partial\theta}Y_{lm}(\theta,\phi) + \dfrac{1}{r}\dfrac{\partial}{\partial r}\left(rj_l(|\lambda_{ql}|r)\right)\dfrac{imY_{lm}(\theta,\phi)}{\sin\theta} \end{pmatrix}$$

where $q$ are non-zeros integers, $l$ are positive integers, $m$ range from $-l$ to $l$, $j_l$ is the $l$-th order spherical Bessel function, $Y_{lm}$ is the normalized spherical harmonic function, $i$ is the imaginary unit. The eigenvalue $\lambda_{ql}$ corresponds to the eigenvalue of the curl operator, and for $q > 0$, $\lambda_{ql} = -\lambda_{-ql}$ and $\lambda_{ql} = Z_{ql}/a$, where $Z_{ql}$ is the $q$-th zero of $j_l(r)$.

The wavenumber $\hat{k}$ can be defined as $|\lambda_{ql}|$. The helical-wave energy spectrum for a flow in a spherical domain is then given by

$$E(k) = \frac{1}{2V} \sum_{|\lambda_{ql}| = k} c_{qlm} c_{qlm}^*.$$

*Channel domain solution.* The solution to Eq. (22) for a channel domain is derived in Cartesian coordinates using the method of separation of variables[41]. The wavenumber vectors are represented as

$$\boldsymbol{k} = (k_x, k_y, k_z) = \left( \frac{2\pi m}{L_x}, \frac{l\pi}{2h}, \frac{2\pi n}{L_z} \right),$$

where $m$ and $n$ as integers, and $l$ is a non-negative integer. The expression for helical-waves is then given by

$$\boldsymbol{B}_k(\boldsymbol{k}, \boldsymbol{x}) = C \cdot \begin{cases} (0,0,1), & \text{for } m = n = 0 \text{ and } l = 0, \\ (s\cos(k_y y), 0, \sin(k_y y)), & \text{for } m = n = 0 \text{ and } l \text{ is odd}, \\ (s\sin(k_y y), 0, -\cos(k_y y)), & \text{for } m = n = 0 \text{ and } l \text{ is even}, \\ \dfrac{1}{k} \begin{pmatrix} -i(k_y^2 + k_z^2)^{1/2} \cos(k_y y - s\alpha) \\ s(k_z^2 + k_x^2)^{1/2} \cos(k_y y) \\ i(k_x^2 + k_y^2)^{1/2} \cos(k_y y + s\beta) \end{pmatrix} \exp[i(k_x x + k_z z)], & \text{for } m^2 + n^2 \neq 0 \text{ and } l \text{ is odd}, \\ \dfrac{1}{k} \begin{pmatrix} -i(k_y^2 + k_z^2)^{1/2} \sin(k_y y - s\alpha) \\ s(k_z^2 + k_x^2)^{1/2} \sin(k_y y) \\ i(k_x^2 + k_y^2)^{1/2} \sin(k_y y + s\beta) \end{pmatrix} \exp[i(k_x x + k_z z)], & \text{for } m^2 + n^2 \neq 0 \text{ and } l \text{ is even}. \end{cases}$$

$$C = \frac{1}{\sqrt{2hL_xL_z}}, \qquad \tan\alpha = \frac{k_x k_y}{k k_z}, \qquad \tan\beta = \frac{k_y k_z}{k k_x}, \qquad k = |\boldsymbol{k}|, \qquad s = \pm 1.$$

The typical wavelength $l_k$ for each helical mode is defined as

$$l_k = \begin{cases} \sqrt{L_x^2 + L_z^2 + (2h)^2}, & \text{for } m = n = 0 \text{ and } l = 0, \\ \sqrt{L_x^2 + L_z^2 + (2h/l)^2}, & \text{for } m = n = 0 \text{ and } l \neq 0, \\ \sqrt{(L_x/(2m))^2 + (L_z/(2n))^2 + (2h/l)^2}, & \text{for } m^2 + n^2 \neq 0 \text{ and } l \neq 0. \end{cases}$$

The helical-wave energy spectrum for a channel domain is defined as

$$E(k) = \frac{1}{2V} \sum_{2\pi/l_k = k} c_{mln} c_{mln}^*. \tag{23}$$

**Cylindrical Annular Domain Solution.** In a TC flow, the cylindrical domain is periodic in $z$-direction with height $H$, and is partially utilized in $\theta$-direction, maintaining a rotational symmetry of order $n_s$. Under periodic boundary conditions in the wall-parallel directions, Eq. (22) is solved in cylindrical coordinates. The helical-waves $\boldsymbol{B}_k = (B_r, B_\theta, B_z)$ in the cylindrical geometry domain of TCF's type are expressed as

$$\boldsymbol{B}_k = (im\lambda_k F/r + 2\pi i n \cdot F'/H, \quad -\lambda_k F' - 2\pi mnF/r/H, \quad \mu^2 F) \exp[i(m\theta + 2\pi nz/H)],$$

where $F(r)$ is given by

$$F(r) = C_1 J_{|m|}(\mu r) + C_2 Y_{|m|}(\mu r),$$

with $\lambda_k^2 - (2n\pi/H)^2 = \mu^2$ (where $\mu > 0$), and $J_l(r)$, $Y_l(r)$ are the first and second kind Bessel functions, respectively. $n$ are integers, $m$ is an integer multiple of $n_{\text{sym}}$. Constants $C_1, C_2$ and the eigenvalue $\lambda_k$ are determined numerically based on the conditions of non-penetration at the walls. This yields

$$C_1 \left( \frac{m\lambda_k}{r_i} J_{|m|}(\mu r_i) + \frac{2\pi \mu n}{H} J_{|m|}{}'(\mu r_i) \right) + C_2 \left( \frac{m\lambda_k}{r_i} Y_{|m|}(\mu r_i) + \frac{2\pi \mu n}{H} Y_{|m|}{}'(\mu r_i) \right) = 0,$$

$$C_1\left(\frac{m\lambda_k}{r_o}J_{|m|}(\mu r_o) + \frac{2\pi\mu n}{H}J_{|m|}{}'(\mu r_o)\right) + C_2\left(\frac{m\lambda_k}{r_o}Y_{|m|}(\mu r_o) + \frac{2\pi\mu n}{H}Y_{|m|}{}'(\mu r_o)\right) = 0,$$

and the normalization of the bases

$$\int_\Omega \boldsymbol{B}_k \cdot \boldsymbol{B}_k^* \, dV = \frac{2\pi H}{n_s}\int_{r_i}^{r_o}[(m\lambda_k F/r)^2 + (\lambda_k F')^2 + (\lambda_k^2 F)^2] \, r \, dr = 1.$$

Similar to the typical wavelength of helical-waves in a channel, the wavelength of each helical wave in the TCF can be expressed as

$$l_k = \begin{cases} \sqrt{H^2 + (2\pi r_i/n_s)^2 + (d/l)^2}, & \text{for } m = n = 0, \\ \sqrt{(\pi r_i/m)^2 + (H/(2n))^2 + (d/l)^2}, & \text{for } m^2 + n^2 \neq 0. \end{cases}$$

Here, $l$ is the index of $|\lambda_k|$ ordered in increasing magnitude. The definitions of wavenumber and helical-wave energy spectrum for the TCF are identical to those for the channel domain (Eq. (23)).

Implementation of a filtered velocity field

In Eq. (21), the helical energy spectrum of the bulk region is derived from helical-wave coefficients obtained by projecting the velocity field onto the helical bases. A bandpass-filtered velocity field $\widehat{\boldsymbol{u_c}}(\boldsymbol{x}, t)$ is reconstructed by selecting helical modes whose wavenumber lie in the range from just below the lowest wavenumber of the inertial range to just below the lowest wavenumber of the far dissipative range of the energy spectrum. This range is denoted by $k \in [k_{l_s}, k_{l_e}]$, so that

$$\widehat{\boldsymbol{u_c}}(\boldsymbol{x}, t) = \sum_{k_{l_s} \leq k \leq k_{l_e}} c_k(t)\boldsymbol{B}_k(\boldsymbol{x}).$$

Velocity in the bulk region

The velocity field of the bulk region, denoted as $\boldsymbol{u}_c(\boldsymbol{x}, t)$, is extracted from the total flow field $\boldsymbol{u}(\boldsymbol{x}, t)$ using a Tukey window function, $w(\boldsymbol{x}, \Delta)$, commonly used in signal processing[47]. The specific parameters for the various BDT cases are provided in Table S5. Within the window, the velocity field is expressed as $\boldsymbol{u}_c(\boldsymbol{x}, t) = w(\boldsymbol{x}, \Delta) \cdot \boldsymbol{u}(\boldsymbol{x}, t)$. While it is theoretically possible to introduce a gauge term $\nabla\chi$ and an additional solenoidal field with arbitrarily small energy to enforce incompressibility and smooth the boundaries of the window, such modifications do not affect the helical-wave energy spectrum. Furthermore, the calculation of structure functions is carried out exclusively within the plateau of the window, ensuring that the transient regions do not influence the results.

Velocity structure functions

In accordance with the definition of the longitudinal velocity structure function, an integral over five-dimensional space is performed for a given spatial distance $\ell$. The procedure combines spherical coordinates and Cartesian coordinates, as expressed by

$$S_p(\ell) = \langle\frac{1}{V_\Omega}\frac{1}{4\pi\ell^2}\int_\Omega\int_0^{2\pi}\int_0^\pi |\Delta u_i e_i|^p \ell^2 \sin\theta \, d\theta \, d\phi \, dx_1 dx_2 dx_3\rangle_t, \tag{24a}$$

$$D_p(\ell) = \langle\frac{1}{V_\Omega}\frac{1}{4\pi\ell^2}\int_\Omega\int_0^{2\pi}\int_0^\pi (\Delta u_i e_i)^p \ell^2 \sin\theta \, d\theta \, d\phi \, dx_1 dx_2 dx_3\rangle_t. \tag{24b}$$

Here, $\Delta u_i = u_i(x_1 + \ell\sin\theta\cos\phi, x_2 + \ell\sin\theta\sin\phi, x_3 + \ell\cos\theta) - u_i(x_1, x_2, x_3)$ represents the velocity increment, and $\boldsymbol{e}_i = \boldsymbol{\ell}_i/\ell = (\sin\theta\cos\phi, \sin\theta\sin\phi, \cos\theta)$ is the unit vector along the direction of separation. The symbol $\langle\cdot\rangle_t$ denotes the time average. To promote the efficiency of the integration of Eqs. S5, we apply the Monte Carlo method by randomly selecting $N_\ell$ point pairs $(\boldsymbol{x}, \boldsymbol{x} + \boldsymbol{\ell})_j (j = 1, 2, 3, \ldots, N_\ell)$ separated by a distance of $\ell$, and then averaging the $p$th power of longitudinal velocity increments of these pairs. The sampled averages for the structure functions, both with and without absolute value, are expressed as

$$S_p(\ell) = \langle\frac{1}{N_\ell}\sum_{j=1}^{N_\ell}\left|\Delta u_i^j e_i^j\right|^p\rangle_t, \qquad D_p(\ell) = \langle\frac{1}{N_\ell}\sum_{j=1}^{N_\ell}\left(\Delta u_i^j e_i^j\right)^p\rangle_t.$$

We propose a GMM (Eq. (12)) for modeling the PDF of $\delta v_{\ell_0}$ within the inertial range. It is non-negative across its domain of definition, and the total area under the curve is equal to 1. This means that $m_1 + m_2 = 1$. And

$$
\begin{aligned}
\int_{-\infty}^{\infty} P(\delta v_\ell) d\delta v_\ell &= \int_{-\infty}^{\infty} \sum_{i=1}^{2} \sum_{k=0}^{\infty} \sum_{k=0}^{\infty} e^{-\lambda} \frac{\lambda^k}{k!} \frac{m_i}{\sqrt{2\pi}\sigma_i(\ell/\ell_0)^\gamma \beta^k} \exp\left[-\frac{(\delta v_l - \mu_i(\ell/\ell_0)^\eta \beta^k)^2}{2\sigma_i^2(\ell/\ell_0)^{2\gamma}\beta^{2k}}\right] d\delta v_\ell \\
&= \sum_{i=1}^{2} \sum_{k=0}^{\infty} e^{-\lambda} \frac{\lambda^k}{k!} m_i \int_{-\infty}^{\infty} \frac{1}{\sqrt{2\pi}\sigma_i(\ell/\ell_0)^\gamma \beta^k} \exp\left[-\frac{(\delta v_l - \mu_i(\ell/\ell_0)^\eta \beta^k)^2}{2\sigma_i^2(\ell/\ell_0)^{2\gamma}\beta^{2k}}\right] d\delta v_\ell \\
&= \sum_{k=0}^{\infty} e^{-\lambda} \frac{\lambda^k}{k!} \cdot \sum_{i=1}^{2} m_i = e^{-\lambda} \sum_{k=0}^{\infty} \frac{\lambda^k}{k!} = 1
\end{aligned}
$$

The $p$th-order moments of $|\delta v_\ell|$ and the odd $p$th-order moments of $\delta v_\ell$ are given by

$$
S_p(\ell) = \int_{-\infty}^{\infty} |\delta v_\ell|^p P(\delta v_\ell) d\delta v_\ell = \Gamma\left(\frac{p+1}{2}\right)\sqrt{\frac{2^p}{\pi}}\left(\frac{\ell}{\ell_0}\right)^{p\gamma + C(1-\beta^p)} \sum_{i=1}^{2} m_i \sigma_i^p \, M\left(\frac{1-p}{2}, \frac{3}{2}, -\frac{\mu_i^2}{2\sigma_i^2}\left(\frac{\ell}{\ell_0}\right)^{2(\eta-\gamma)}\right),
$$

$$
D_p(\ell) = \int_{-\infty}^{\infty} \delta v_\ell^p P(\delta v_\ell) d\delta v_\ell = \Gamma\left(\frac{p+2}{2}\right)\sqrt{\frac{2^{p+1}}{\pi}}\left(\frac{\ell}{\ell_0}\right)^{(p-1)\gamma + \eta + C(1-\beta^p)} \sum_{i=1}^{2} m_i \mu_i \sigma_i^{p-1} \, M\left(\frac{1-p}{2}, \frac{3}{2}, -\frac{\mu_i^2}{2\sigma_i^2}\left(\frac{\ell}{\ell_0}\right)^{2(\eta-\gamma)}\right),
$$

where $M(a,b,z)$ denotes Kummer's confluent hypergeometric function and $\Gamma(z)$ represents the Gamma function. The analysis indicates that $|\mu_i/\sigma_i| \ll (\ell_0/\ell)^{\eta-\gamma}$ and $|\mu_i/\sigma_i| \ll 1$ (Table S3), which allows the confluent hypergeometric functions to the approximated as 1. Hence, it can be derived

$$
S_p(\ell) = \Gamma\left(\frac{p+1}{2}\right)\sqrt{2^p/\pi}\left(\frac{\ell}{\ell_0}\right)^{p\gamma + C(1-\beta^p)} \sum_{i=1}^{2} m_i \sigma_i^p \sim \ell^{p\gamma + C(1-\beta^p)},
$$

$$
D_p(\ell) = \Gamma\left(\frac{p+2}{2}\right)\sqrt{2^{p+1}/\pi}\left(\frac{\ell}{\ell_0}\right)^{(p-1)\gamma + \eta + C(1-\beta^p)} \sum_{i=1}^{2} m_i \mu_i \sigma_i^{p-1} \sim \ell^{(p-1)\gamma + \eta + C(1-\beta^p)}.
$$

Local isotropy necessitates that $D_1 = 0$, which implies $\sum_{i=1}^{2} m_i \mu_i = 0$. Both experiments and numerical simulations reveals that $D_3 < 0$, indicating that $\sum_{i=1}^{2} m_i \mu_i \sigma_i^2 = m_1 \mu_1(\sigma_1^2 - \sigma_2^2) = m_1 \mu_1(\sigma_1 - \sigma_2)(\sigma_1 + \sigma_2) < 0$. These findings, combined with the fact that both $\sigma_1$ and $\sigma_2$ are greater than zero, lead to the conclusion that

$$
\sum_{i=1}^{2} m_i \mu_i \sigma_i^{p-1} \xrightarrow{p-1=2q} m_1 \mu_1(\sigma_1^{2q} - \sigma_2^{2q}) = m_1 \mu_1(\sigma_1 - \sigma_2)(\sigma_1^q + \sigma_2^q)\sum_{j=1}^{q} \sigma_1^{q-j}\sigma_2^{j-1} < 0.
$$

Therefore, $D_p < 0$ for all odd $p > 1$.

# Data availability

The data generated in this study are available from the corresponding author upon reasonable request.

# Code availability

The code used for numerical simulations, data processing, and figure generation are available from the corresponding author upon reasonable request.

# Supplementary Material: Universal scaling laws of boundary-driven turbulence

Yong-Ying Zeng[1], Zi-Ju Liao[2], Jun-Yi Li[1,3], Wei-Dong Su[1,*]

[1]State Key Laboratory for Turbulence and Complex Systems and Department of Mechanics and Engineering Science, College of Engineering, Peking University, Beijing 100871, China

[2]Department of Mathematics, College of Information Science and Technology, Jinan University, Guangzhou 510632, China

[3]New Cornerstone Science Laboratory, Center for Combustion Energy, Key Laboratory for Thermal Science and Power Engineering of Ministry of Education, Department of Energy and Power Engineering, Tsinghua University, Beijing 100084, China

[*]Email: swd@pku.edu.cn

There are five figures (Fig. S1, S2, S3, S4, S5) and six tables (Table S1, S2a, S2b, S3, S4, S5) supplemented here.

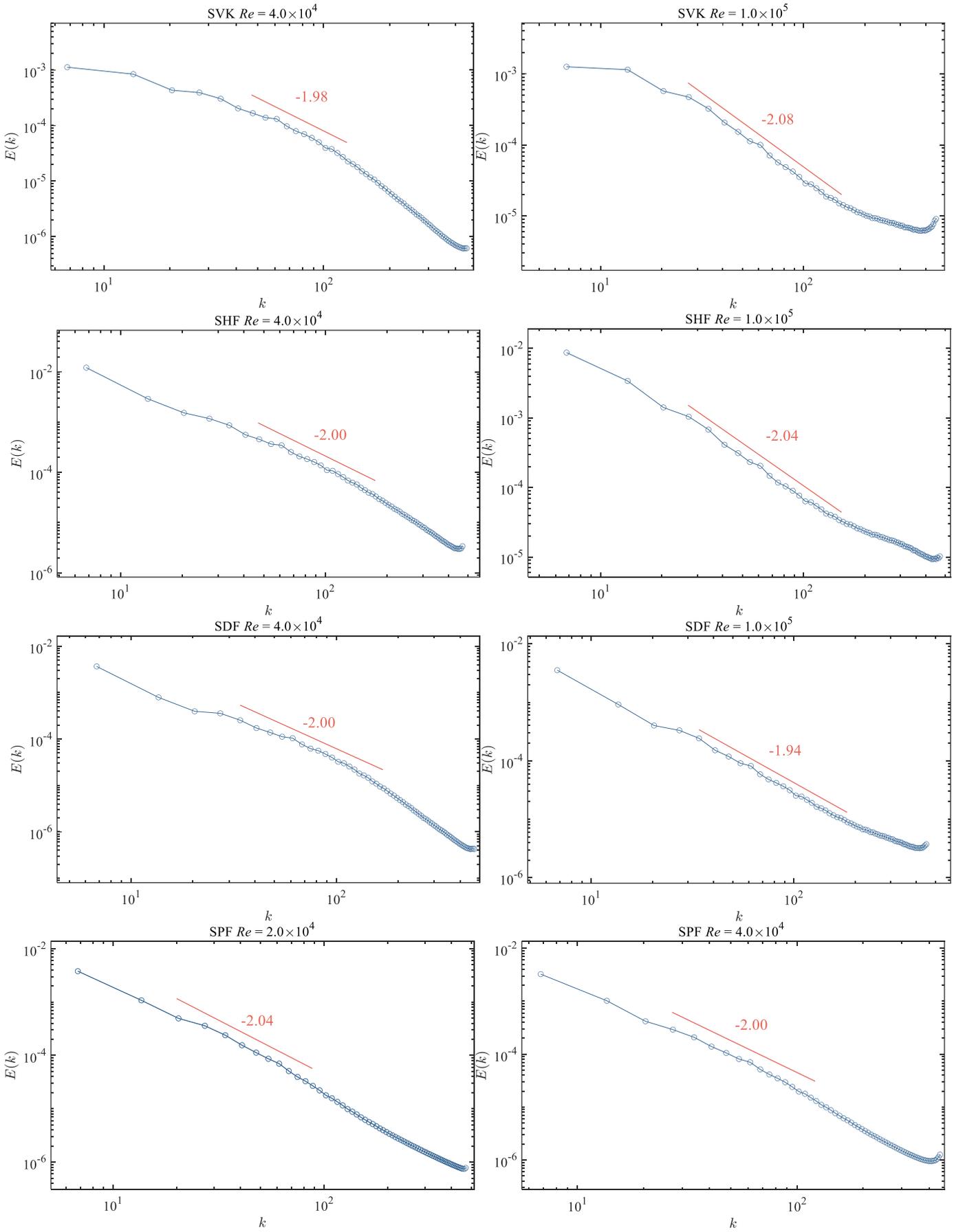

**Fig. S1. Fourier-based energy spectrum in spherical-wall bounded flows.** The time-averaged energy spectrum $E(k)$ was computed in a cubic ($D \times D \times D$) box within the core region, with $D = 0.9238$. The inertial range exhibiting the scaling law $E(k) \sim k^{-\alpha}$ is demarcated by red solid lines in each panel. All cases demonstrate consistent scaling exponents with $\alpha$ approaching 2 across configurations.

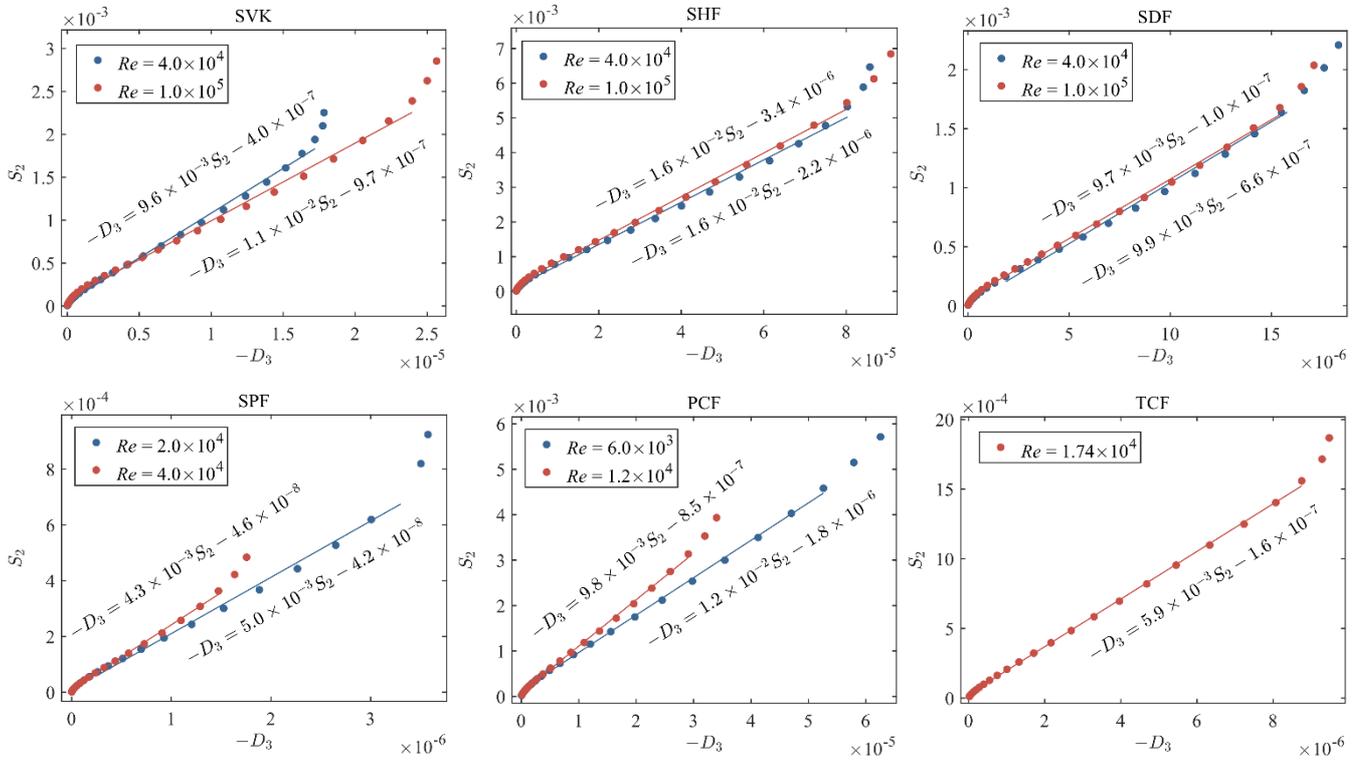

**Fig. S2. The relationship between $S_2(\ell)$ and $D_3(\ell)$ in the inertial range.** The second-order velocity structure function $S_2(\ell)$ as a function of the negative third-order velocity structure function $-D_3(\ell)$. DNS results (symbols) and the best fits (lines) of $-D_3 = a\,S_2 + b$ within the inertial range (with all prefactors) are shown in figures. The slope $a$ represents the characteristic speed $U$.

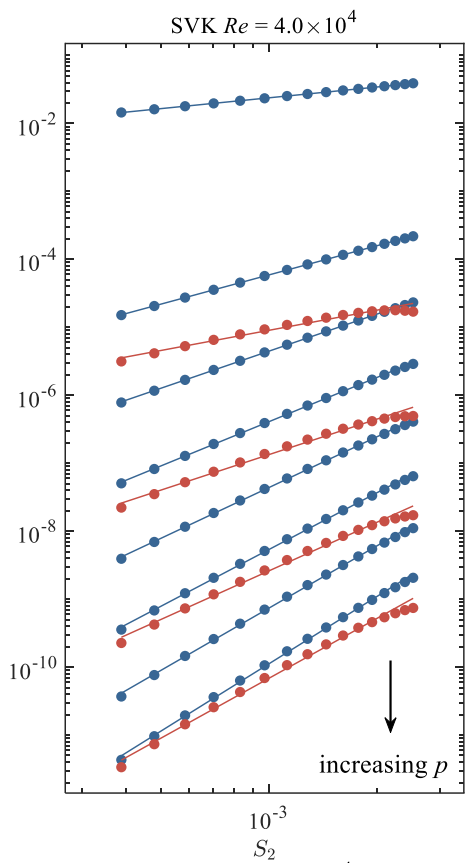

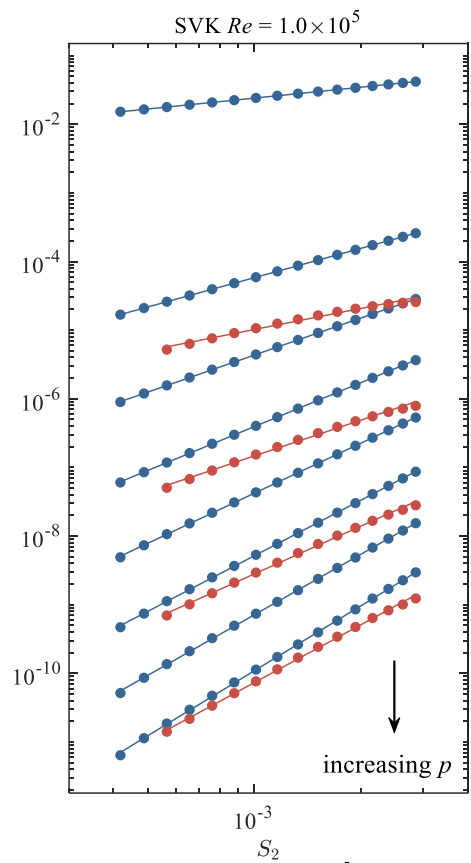

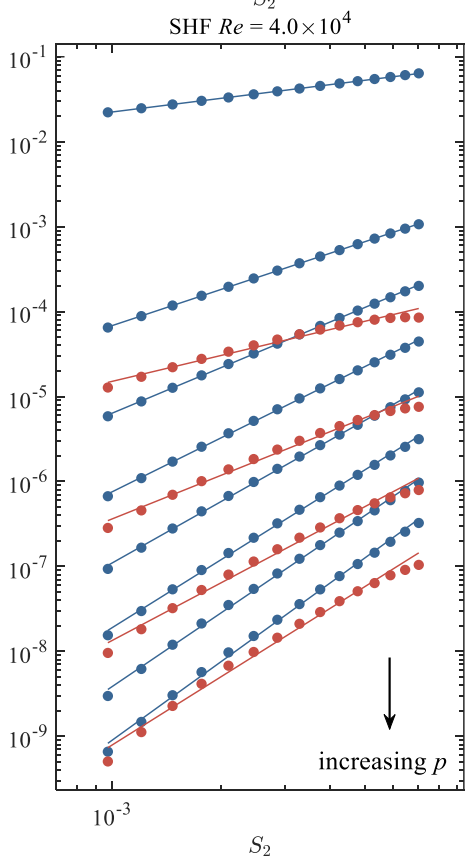

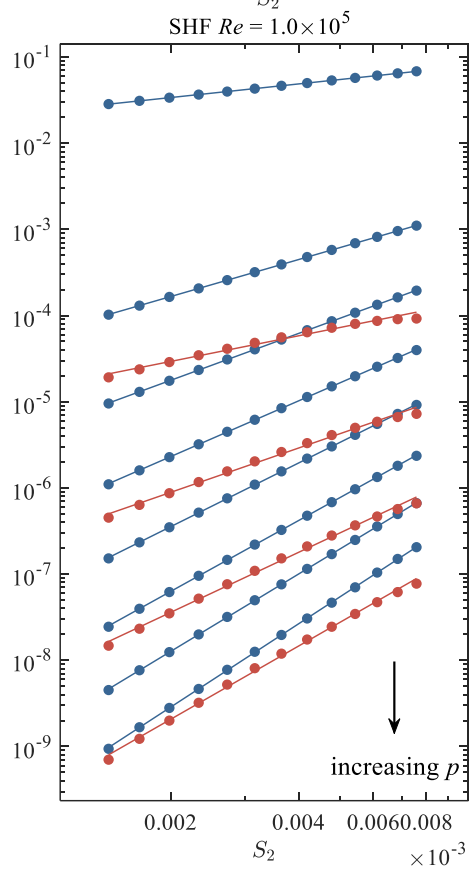

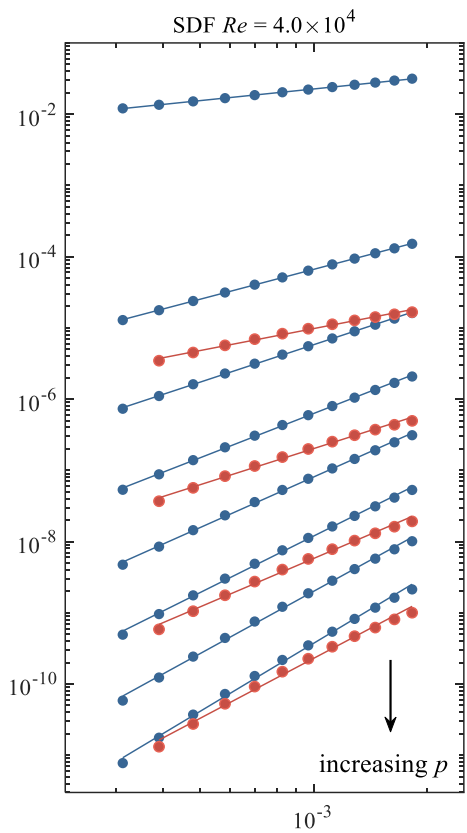

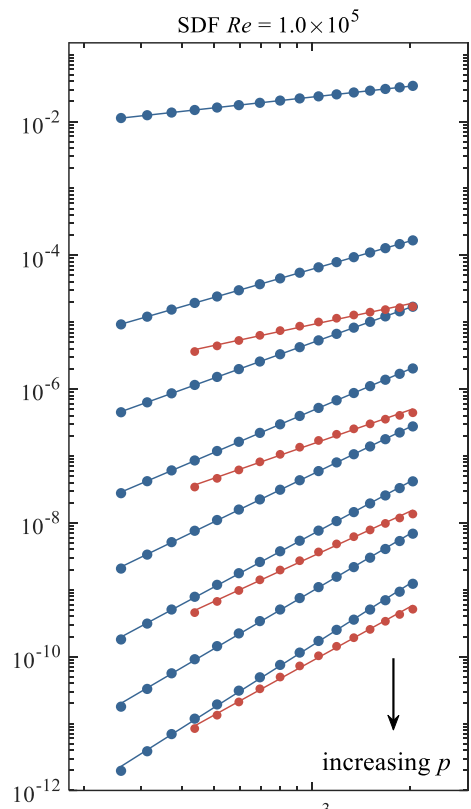

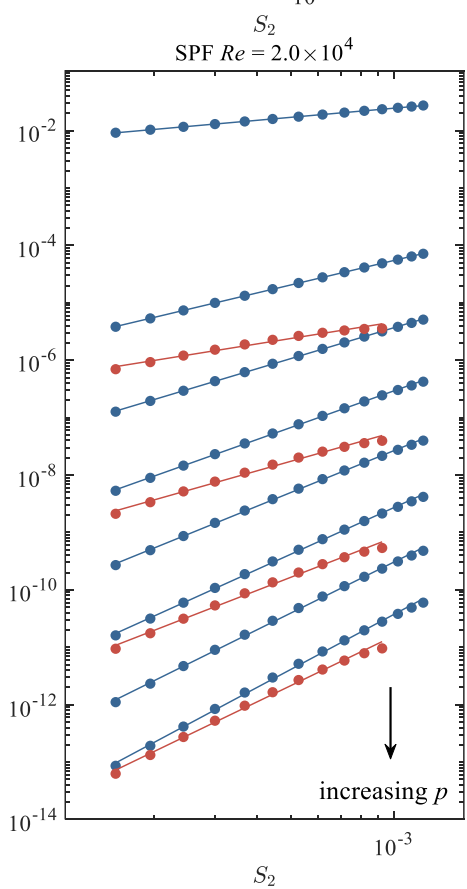

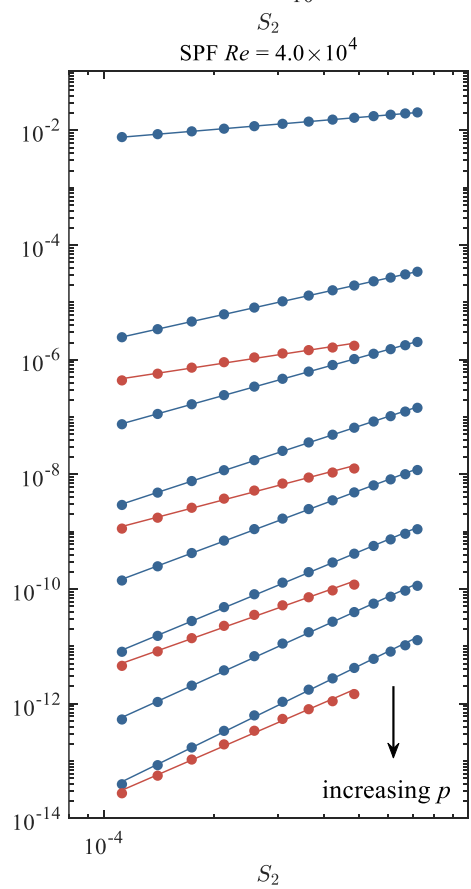

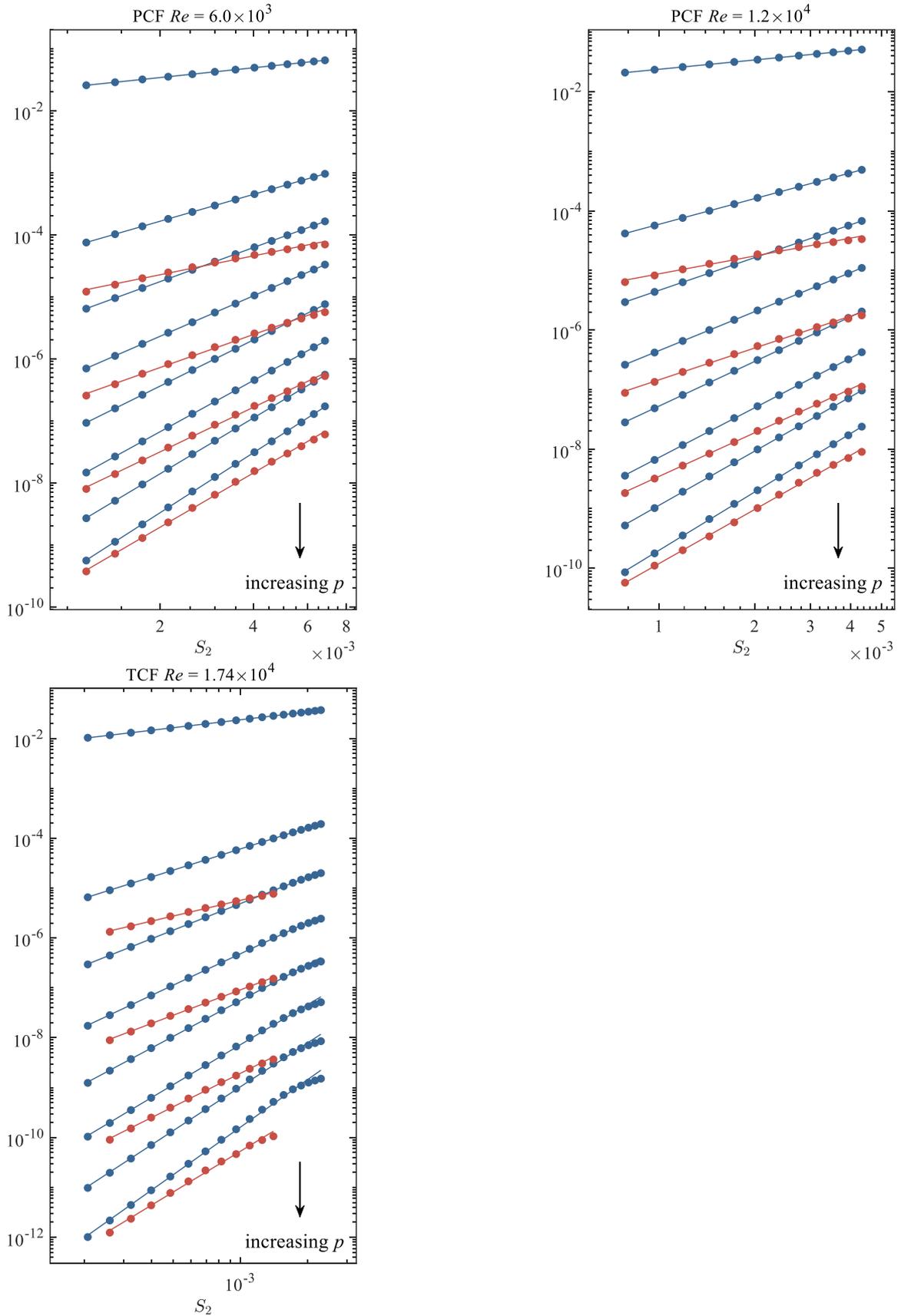

**Fig. S3. Extended self-similarity on velocity structure functions of the filtered velocity field.** DNS (symbols) and fits (solid lines) are illustrated. Velocity structure functions regarding the velocity increment ($D_p(\ell)$, red) and its absolute value ($S_p(\ell)$, blue) versus the second-order velocity structure functions $S_2(\ell)$ are displayed with $p = 1,2,\ldots,9$ from top to bottom. Table S2A offers slopes for linear least-squares fits. The shaded regions in Fig. 2 and Fig. 3 in the main text correspond to the length scale ranges of the objects (bule) depicted above. The transform relation between $\ell$ in physical space and $k$ in spectral space is $\ell \cdot k = $ constant. For cylindrical and channel domains, this constant is set to 1, while for the spherical domain, it is set to 2.9.

**Table S1. Characteristic speed $U$, projectile speed $U_p$, and their ratio in all of the cases.**

| $Re(\times 10^4)$ | 10.0 | 4.0 | 10.0 | 4.0 | 10.0 | 4.0 | 4.0 | 2.0 | 1.2 | 0.6 | 1.74 |
|---|---|---|---|---|---|---|---|---|---|---|---|
| | SVK | | SHF | | SDF | | SPF | | PCF | | TCF |
| $U^{\mathrm{a}}$ $(\times 10^{-3})$ | 11.0 | 9.6 | 16.0 | 16.0 | 9.7 | 9.9 | 4.3 | 5.0 | 9.8 | 12.0 | 5.9 |
| $U_p^{\mathrm{b}}$ $(\times 10^{-2})$ | 3.6 | 3.4 | 5.3 | 5.6 | 3.3 | 2.8 | 3.8 | 4.4 | 4.6 | 6.6 | 4.2 |
| $U_p/U$ | 3.3 | 3.6 | 3.3 | 3.5 | 3.4 | 2.1 | 8.9 | 8.8 | 4.7 | 5.5 | 7.2 |

[a] Characteristic speed defined as $-D_3(\ell)/S_2(\ell)$, which is independent of $\ell$ within the inertial range
[b] Projectile speed defined as the normal velocity averaged over the interface between the bulk and the boundary layer

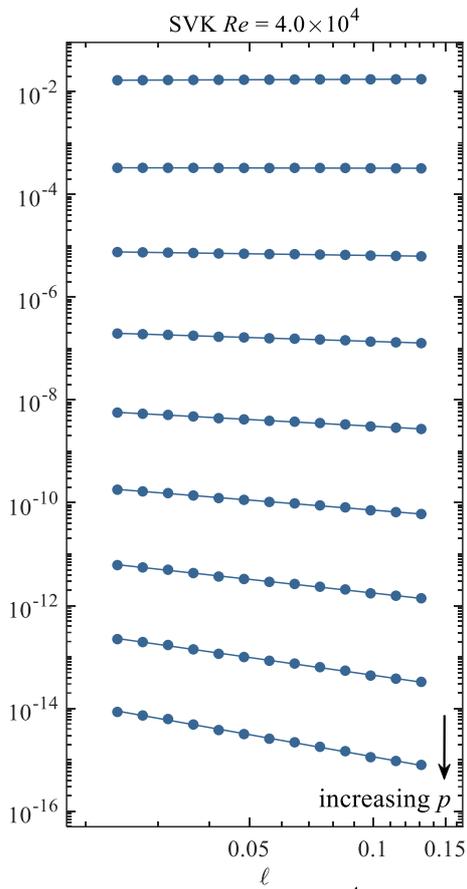

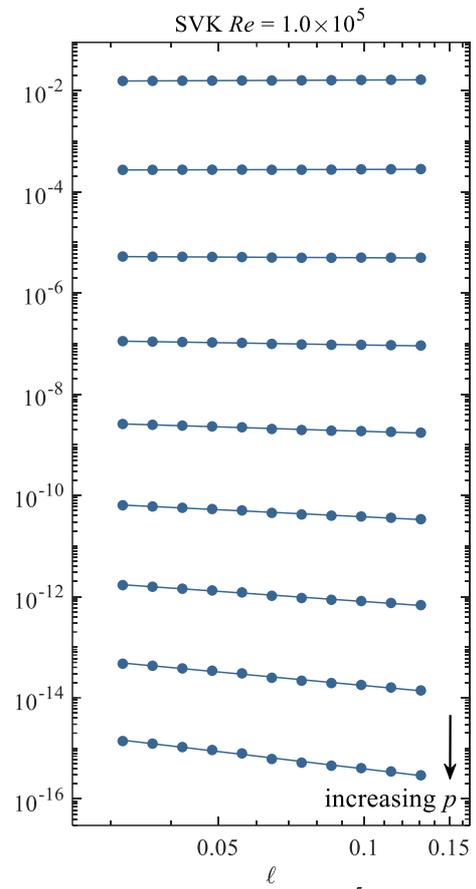

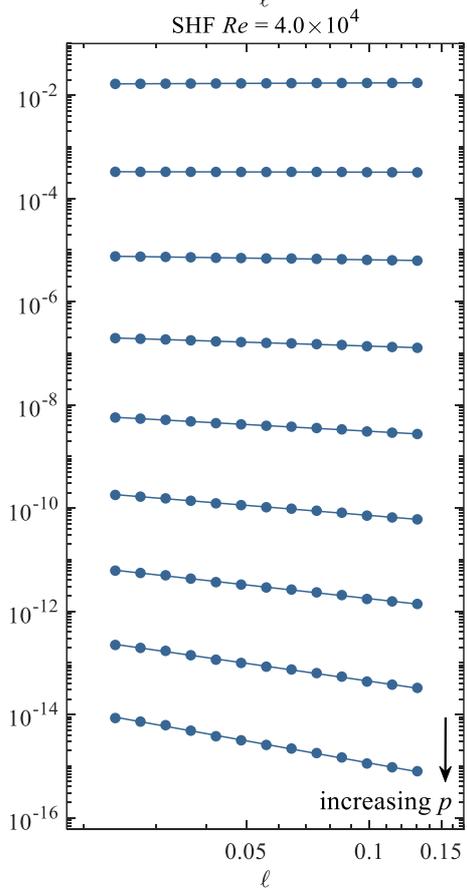

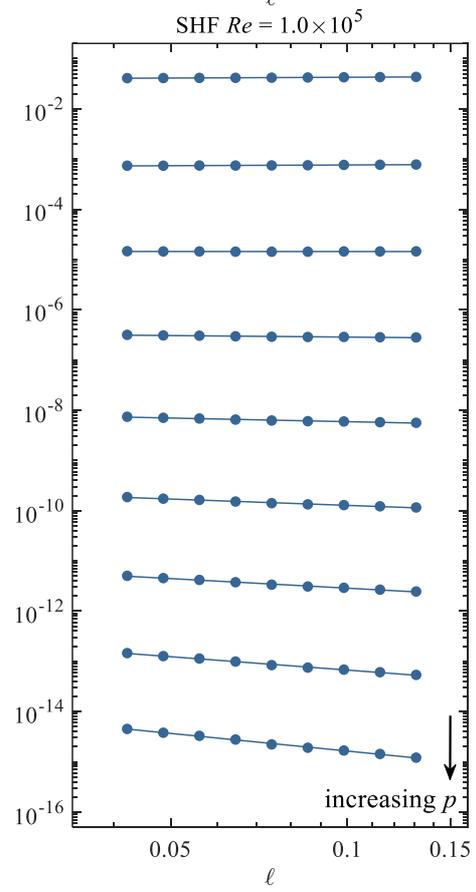

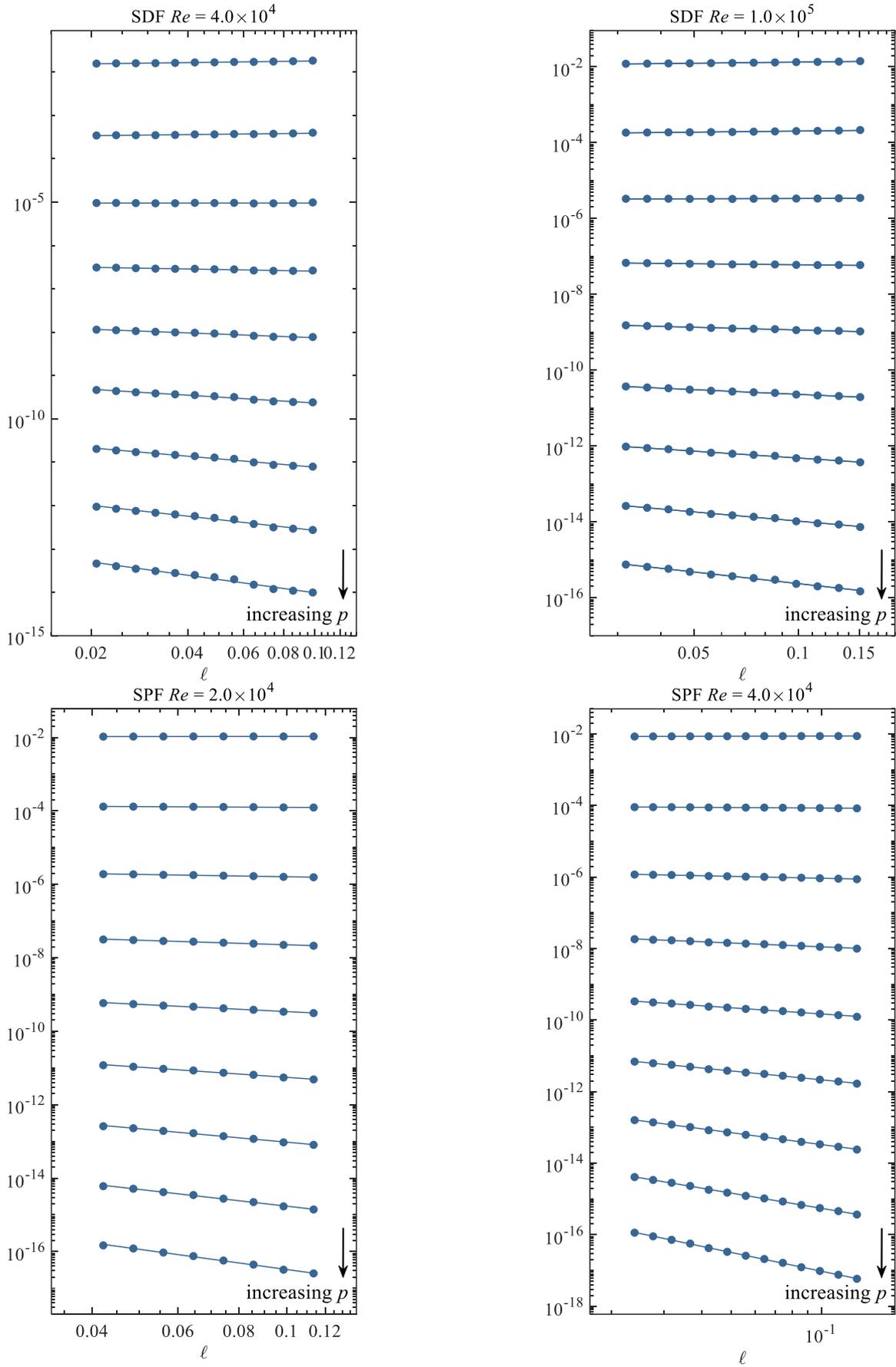

**Fig. S4. Scaling behavior of the coarse-grained energy dissipation rate for the filtered velocity field.** DNS (symbols) and fits (solid lines) are illustrated in the log-log plot. The coarse-grained energy dissipation rate $\epsilon_\ell$ is fitted as a power law with respect to spatial scale $\ell$ within the inertial range, i.e., $\langle \epsilon_\ell^p \rangle \sim \ell^{\tau_p}$, with $p$ taking values from 1/2, 1, 3/2, ..., 9/2. The scaling exponents $\tau_p$ are illustrated in Fig. 6A in the main text.

**Table S2A. Scaling exponents of velocity structure functions calculated by extended self-similarity for all of the cases.**

| $Re(\times 10^4)$ | 10.0 | 4.0 | 10.0 | 4.0 | 10.0 | 4.0 | 4.0 | 2.0 | 1.2 | 0.6 | 1.74 | 5.4[a] | 6.9[a] | | |
|---|---|---|---|---|---|---|---|---|---|---|---|---|---|---|---|
| $p$ | SVK | | SHF | | SDF | | SPF | | PCF | | | TCF | | $\zeta_{p,BDT}$ | $\xi_{p,BDT}$ |
| 1 | 0.53 | 0.53 | 0.53 | 0.53 | 0.54 | 0.54 | 0.53 | 0.53 | 0.52 | 0.52 | 0.53 | 0.53 | 0.53 | 0.53 | |
| 2 | 1 | 1 | 1 | 1 | 1 | 1 | 1 | 1 | 1 | 1 | 1 | 1 | 1 | 1 | 1 |
| 3 | 1.42 | 1.42 | 1.43 | 1.41 | 1.40 | 1.40 | 1.42 | 1.42 | 1.44 | 1.43 | 1.41 | 1.43 | 1.43 | 1.41 | |
| 3* | 1.02 | 0.98 | 0.99 | 1.01 | 1.04 | 1.03 | 0.97 | 0.96 | 1.00 | 1.01 | 1.05 | | | | 1 |
| 4 | 1.79 | 1.81 | 1.81 | 1.78 | 1.76 | 1.75 | 1.77 | 1.79 | 1.83 | 1.82 | 1.77 | 1.81 | 1.81 | 1.78 | |
| 5 | 2.13 | 2.16 | 2.15 | 2.11 | 2.08 | 2.07 | 2.09 | 2.12 | 2.18 | 2.16 | 2.09 | 2.14 | 2.16 | 2.11 | |
| 5* | 1.73 | 1.73 | 1.71 | 1.70 | 1.70 | 1.69 | 1.67 | 1.69 | 1.78 | 1.76 | 1.69 | | | | 1.70 |
| 6 | 2.43 | 2.48 | 2.46 | 2.40 | 2.36 | 2.36 | 2.38 | 2.42 | 2.49 | 2.47 | 2.39 | 2.46 | 2.46 | 2.41 | |
| 7 | 2.69 | 2.77 | 2.73 | 2.65 | 2.62 | 2.64 | 2.64 | 2.70 | 2.78 | 2.74 | 2.66 | 2.71 | 2.74 | 2.68 | |
| 7* | 2.32 | 2.37 | 2.31 | 2.25 | 2.24 | 2.26 | 2.25 | 2.32 | 2.44 | 2.38 | 2.23 | | | | 2.27 |
| 8 | 2.93 | 3.04 | 2.98 | 2.88 | 2.86 | 2.91 | 2.88 | 2.95 | 3.03 | 2.98 | 2.92 | 2.94 | 2.99 | 2.94 | |
| 9 | 3.15 | 3.28 | 3.21 | 3.07 | 3.09 | 3.18 | 3.11 | 3.19 | 3.26 | 3.20 | 3.16 | 3.13 | 3.23 | 3.18 | |
| 9* | 2.80 | 2.92 | 2.82 | 2.67 | 2.70 | 2.80 | 2.76 | 2.88 | 3.02 | 2.90 | 2.74 | | | | 2.79 |

[a] From experimental data as time series [1].

* Scaling exponents of $\langle (\delta v_\ell)^p \rangle$ for odd $p$.

**Table S2B. Relative error (%) of scaling exponents between results presented in Table S2A and the theoretical predictions.**

| $Re(\times 10^4)$ | 10.0 | 4.0 | 10.0 | 4.0 | 10.0 | 4.0 | 4.0 | 2.0 | 1.2 | 0.6 | 1.74 | $5.4^a$ | $6.9^a$ |
|---|---|---|---|---|---|---|---|---|---|---|---|---|---|
| $p$ | SVK | | SHF | | SDF | | SPF | | PCF | | | TCF | |
| 1 | 1.11 | 1.16 | 1.61 | 0.02 | 0.76 | 2.01 | 0.43 | 0.93 | 2.39 | 2.20 | 0.95 | 0.96 | 0.95 |
| 2 | 0 | 0 | 0 | 0 | 0 | 0 | 0 | 0 | 0 | 0 | 0 | 0 | 0 |
| 3 | 0.97 | 0.97 | 1.04 | 0.05 | 0.55 | 1.12 | 0.02 | 0.44 | 1.74 | 1.48 | 0.05 | 1.22 | 1.22 |
| $3^*$ | 2.25 | 1.59 | 1.18 | 1.12 | 3.98 | 2.53 | 3.00 | 4.10 | 0.36 | 0.74 | 4.75 | | |
| 4 | 1.79 | 1.76 | 1.65 | 0.08 | 1.01 | 1.67 | 0.30 | 0.62 | 2.87 | 2.31 | 0.46 | 2.05 | 2.05 |
| 5 | 0.94 | 2.41 | 1.94 | 0.02 | 1.46 | 1.92 | 0.70 | 0.68 | 3.47 | 2.56 | 0.79 | 1.67 | 2.35 |
| $5^*$ | 2.20 | 2.12 | 0.74 | 0.36 | 0.27 | 0.51 | 1.55 | 0.59 | 3.92 | 3.98 | 0.38 | | |
| 6 | 0.75 | 2.89 | 2.00 | 0.37 | 1.90 | 1.94 | 1.14 | 0.68 | 3.63 | 2.46 | 0.92 | 2.07 | 2.06 |
| 7 | 0.36 | 3.22 | 1.87 | 1.04 | 2.31 | 1.68 | 1.64 | 0.62 | 3.45 | 2.04 | 0.86 | 1.17 | 2.23 |
| $7^*$ | 2.05 | 4.44 | 1.90 | 0.83 | 1.23 | 0.47 | 0.95 | 2.07 | 7.38 | 4.91 | 1.96 | | |
| 8 | 0.20 | 3.38 | 1.56 | 2.08 | 2.65 | 1.07 | 1.85 | 0.49 | 3.05 | 1.45 | 0.64 | 0.16 | 1.61 |
| 9 | 0.99 | 3.34 | 1.09 | 3.49 | 2.90 | 0.05 | 2.01 | 0.25 | 2.52 | 0.80 | 0.40 | 1.53 | 1.61 |
| $9^*$ | 1.27 | 5.66 | 2.09 | 3.63 | 2.53 | 1.23 | 0.22 | 4.04 | 9.25 | 4.75 | 1.95 | | |

ᵃ From experimental data as time series [1].

* The inferred value in the calculation of relative error for $\xi_{p,BDT}$.

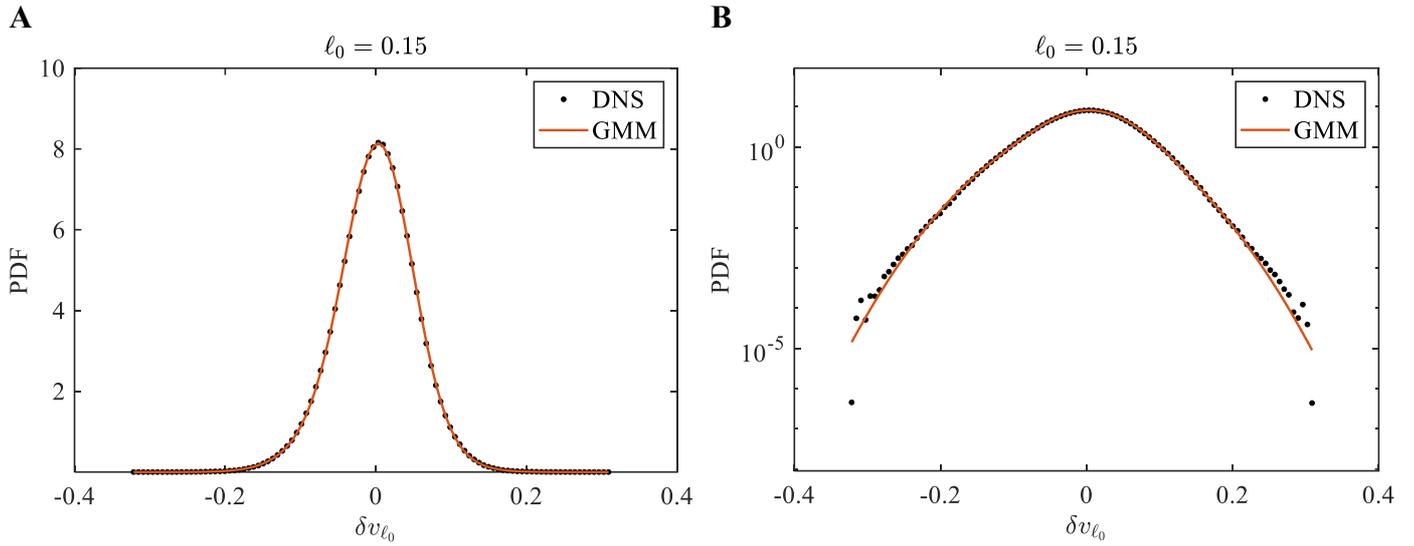

**Fig. S5. PDF of velocity increments at the largest scale of the inertial range.** The case of SVK is shown as a representative example. DNS (symbols) and fits (red solid lines) are illustrated in **A,** Linear plot and **B,** Semi-log plot. The largest scale $\ell_0$ in the inertial range is $0.15$ at $Re = 1.0 \times 10^5$ (Fig. 3). Points calculated from DNS data are fitted with the linear superposition of two Gaussian functions using the EM algorithm. The parameters for fitting are illustrated in Table S3.

**Table S3. Model parameters for the PDF of $\boldsymbol{\delta v_\ell}$.** SVK case at $Re = 1.0 \times 10^5$. $\ell = \ell_0 = 0.15$ is the largest scale within the inertial range. The model is based on the superposition of two Gaussians, where $\mu_i$, $\sigma_i$ and $m_i$ are the mean, the variance and the weight of each Gaussian, respectively.

| $i$ | $m_i$ | $\mu_i$ | $\sigma_i$ | $|\mu_i/\sigma_i|$ |
|-----|-------|---------|------------|--------------------|
| 1 | 0.3687 | $-0.0094$ | 0.0637 | 0.1476 |
| 2 | 0.6313 | 0.0055 | 0.0428 | 0.1285 |

**Table S4. Parameters used for the DNS in the spherical domain.**

|  | SVK | SHF | SDF | SPF |
|---|---|---|---|---|
| *Re* | $1 \times 10^5, 4 \times 10^4$ | | | $4 \times 10^4, 2 \times 10^4$ |

**Table S5. Parameters of the window function for BDT cases.**

| Geometry | Case | Bulk width (radius, radial length, height) | Width of the window ($\Delta$) |
|---|---|---|---|
| Sphere | SVK, SHF, SDF, SPF | 0.80 | 0.20 |
| Cylinder | TCF | 0.35 | 0.07 |
| Channel | PCF | 0.35 | 0.07 |